\documentclass[prd,twocolumn,amsmath,amssymb,floatfix,superscriptaddress,nofootinbib]{revtex4-1}

\usepackage{graphicx}
\usepackage{scrextend}
\usepackage{amssymb}
\usepackage{amsmath}
\usepackage{bm}
\usepackage{color}
\usepackage{multirow}
\usepackage{cprotect}

\DeclareMathAlphabet\mathbfcal{OMS}{cmsy}{b}{n}
\definecolor{darkgreen}{RGB}{50,150,0}
\definecolor{purple}{cmyk}{0.5,1.0,0,0}

\def\edth{\;\raise1.0pt\hbox{$'$}\hskip-6pt\partial}
\def\baredth{\;\overline{\raise1.0pt\hbox{$'$}\hskip-6pt
\partial}}

\newcommand{\Cov}{\mathrm{Cov}}
\newcommand{\PP}{{\phi\phi}}

\def\be{\begin{equation}}
\def\ee{\end{equation}}
\def\ben{\begin{equation} \nonumber}
\def\een{\end{equation}}
\def\ban{\begin{eqnarray*}}
\def\ean{\end{eqnarray*}}
\def\ba{\begin{eqnarray}}
\def\ea{\end{eqnarray}}
\def\({\left(}
\def\){\right)}

\newcommand{\Comment}[1]{{}}

\definecolor{ultramarine}{rgb}{0.07, 0.04, 0.56}
\definecolor{cadmiumgreen}{rgb}{0.0, 0.42, 0.24}
\definecolor{indigo(dye)}{rgb}{0.0, 0.25, 0.42}
\usepackage[linktocpage=true]{hyperref}
\hypersetup{
colorlinks=true,
citecolor=ultramarine,
linkcolor=cadmiumgreen,
urlcolor=indigo(dye),
pdfauthor={},
pdftitle={},
pdfsubject={}
}

\begin{document}

\title{
Lens covariance effects on likelihood analyses of  CMB power spectra
}

\author{Pavel Motloch}
\affiliation{Kavli Institute for Cosmological Physics,  Enrico Fermi Institute, University of Chicago, Chicago, Illinois 60637, U.S.A}
\affiliation{Department of Physics, University of Chicago, Chicago, Illinois 60637, U.S.A}

\author{Wayne Hu}
\affiliation{Kavli Institute for Cosmological Physics,  Enrico Fermi Institute, University of Chicago, Chicago, Illinois 60637, U.S.A}
\affiliation{Department of Astronomy \& Astrophysics, University of Chicago, Chicago, Illinois 60637, U.S.A}

\begin{abstract}
\noindent

Non-Gaussian correlations induced in   CMB  power spectra by gravitational
lensing {must} be included in likelihood analyses for future CMB experiments.
We present a simple but accurate likelihood model which includes these correlations and use it for
Markov Chain Monte Carlo parameter estimation from simulated lensed CMB maps in the context of
$\Lambda$CDM and extensions which include the sum of neutrino masses or  the
dark energy equation of state $w$. If
lensing-induced covariance is not taken into account  for a CMB-S4 type experiment, the errors for one combination of parameters in each case would be underestimated by more then a
factor of two and lower limits on $w$ could be misestimated substantially.  
The frequency of falsely ruling out the true model or finding tension with other data sets would 
also substantially increase.   {Our} analysis
 also enables a separation of
lens and unlensed information from CMB power spectra, which provides for consistency
tests of the model and, if combined with other such measurements,
a nearly lens-sample-variance free test for systematics and new physics
in the unlensed spectrum.   This  parameterization also leads to a simple effective likelihood
that  can be used to assist model building in case consistency tests of $\Lambda$CDM fail.
\end{abstract}

\maketitle

\section{Introduction}
\label{sec:intro}

Measurements of the cosmic microwave background (CMB) have been instrumental in our
understanding of the composition and evolution of the Universe (see e.g.\ \cite{Ade:2015xua}).
Starting with the initial detection through cross correlation of the quadratic reconstruction of the lensing potential \cite{Hu:2001fa} with galaxy surveys
\cite{Smith:2007rg}, gravitational lensing of the CMB by the large
scale structure of the Universe (see \cite{Lewis:2006fu} for a review) has emerged as a
new source of cosmological information \cite{Hirata:2008cb,Hanson:2013hsb,Das:2011ak,
Keisler:2011aw, Ade:2013tyw,Keisler:2015hfa,Ade:2015zua, Array:2016afx, Sherwin:2016tyf}.
For currently operating and future CMB experiments
\cite{Benson:2014qhw, Henderson:2015nzj, Wu:2016hul, Abazajian:2016yjj},
gravitational lensing will be instrumental in breaking parameter degeneracies affecting
low redshift physics, such as sum of the neutrino masses and properties of the dark
energy. 

On the other hand, lensing  also acts as a source of  lens-sample-variance noise for the detection of
inflationary gravitational waves and light relic particles from CMB power spectra  (see e.g.~\cite{Abazajian:2016yjj}).
Lens sample variance can be removed by reconstructing the lensing potential assuming
that the power spectra and higher point moments are lensed by the same potential.
Recently, Ref.~\cite{Motloch:2016zsl} proposed a consistency test between the two that can 
be used to protect against systematic errors and incorrect assumptions on either side. 

{The covariance between CMB observables induced by lensing also complicates their 
analysis. This covariance was first studied between the lensed temperature and
polarization power spectra \cite{Smith:2006nk, BenoitLevy:2012va}, yet all analyses of CMB experiments
to date have omitted it in the likelihood function (e.g.\ \cite{Ade:2015xua}).
{Once polarization measurements approach the sample variance limit, this covariance
must be included in the analysis \cite{BenoitLevy:2012va} as well as the correlation 
between power spectra and the reconstructed lensing potential
\cite{Green:2016cjr, Peloton:2016kbw, Motloch:2016zsl}.}

 Most previous studies have employed
Fisher forecasts to estimate lens-covariance effects  on  parameter
constraints rather than  a direct likelihood exploration, with the exception of \cite{Peloton:2016kbw} who briefly investigate their impact on neutrino mass constraints in $\Lambda$CDM.
The main goal of this work is filling this gap with a more comprehensive
study of an analysis pipeline from simulated lensed CMB maps to parameter posterior
probabilities, including the principal component parameterization of the lensing
potential introduced in Ref.~\cite{Motloch:2016zsl}.  

{This paper is organized as follows. In \S~\ref{sec:analysis_details} we
present our model
for the likelihood function of full-sky  CMB power spectra data which accounts for
lensing-induced covariance between multipole moments.  We briefly review the analytical model 
for the lensing-induced covariance upon which it is {based \cite{BenoitLevy:2012va}
and} define the
fiducial cosmology and experimental configuration investigated in this work. 
Then in
\S~\ref{sec:cosmological_parameters} we look at effects the non-Gaussian
covariance has on cosmological parameter constraints and point out that 
neglecting this covariance could have a
possibly serious impact on concordance studies. In \S~\ref{sec:lens_pc}
we introduce a
principal component parameterization of lensed CMB power spectra, describe results of our MCMC
analyses of properties of these parameters and use them to explain the effects of
lensing-induced covariance on cosmological parameters. In the same section we then detail
how measurements of these effective parameters
can be used to form consistency tests for the data and assist model building.

In the Appendices we give more details on how we modify code Lenspix which simulates
lensed CMB data to achieve high accuracy (Appendix~\ref{sec:lenspix}){, how effects of
the non-Gaussian covariance change for an experiment which does not measure information on
the largest scales (Appendix~\ref{sec:no_low_ell})} and how we determined
necessary number of lensing principal components
(Appendix~\ref{sec:n_pca}).

\section{Analysis details}
\label{sec:analysis_details}

In this section we first describe our fiducial cosmology and experimental configuration.
Then we comment on analytic model used for the covariance of lensed CMB power spectra,
details of our CMB simulations and Markov Chain Monte Carlo analysis and in a
separate section we describe our model for the data likelihood.

\subsection{Fiducial cosmology and experimental setup}

In this work we investigate a simplified experimental setup of a full sky experiment with
specifications inspired by  CMB Stage 4 \cite{Abazajian:2016yjj}.
Throughout the paper we use capital letters $X, Y, W, Z$ to represent either CMB
temperature or polarization field, i.e.~an element from $\{T, E, B\}$.  In a given
cosmological model, we also abbreviate the set of all cosmological parameters as
$\theta_A$.  We consider the information on $\theta_A$ provided by the CMB power
spectra $C_\ell^{XY}$.

For the fiducial cosmology we take a flat 6 parameter $\Lambda$CDM
model with minimal neutrino mass. For the $\Lambda$CDM
parameters we take $\omega_b = \Omega_b h^2$, the physical baryon density;
$\omega_c = \Omega_c h^2$, the physical cold dark matter density; $n_s$, the tilt of the
scalar power spectrum; $A_s$, its amplitude; and $\tau$, the optical depth to
recombination.  We choose $\theta_*$, the angular scale of the sound horizon
at recombination, as opposed to the
Hubble constant $h$, as the sixth independent parameter.
We also assume that tensor modes are negligible so that
there is no unlensed $B$ mode. Values of the cosmological  parameters for
the fiducial model used in this work are summarized in Table~\ref{tab:fiducial}. 

\begin{table}
\caption{Fiducial parameters used in the analysis} 
\label{tab:fiducial}
\begin{tabular}{c|c}
\hline\hline
Parameter & Fiducial value\\
\hline
$h$ & 0.675\\
$\Omega_c h^2$ & 0.1197 \\
$\Omega_b h^2$ & 0.0222 \\
$n_s$ & 0.9655 \\
$A_s$ & $2.196 \times 10^{-9}$ \\
$\tau$ & 0.06 \\
$\sum m_\nu$ & 60 meV \\
\hline\hline
\end{tabular}
\end{table}

 For noise in temperature
and polarizations, we assume a  noise spectra  \cite{Knox:1995dq}
\be
\label{Gauss_noise}
	N_\ell^{XY} = \Delta_{XY}^2 e^{\ell(\ell+1){\theta_\mathrm{FWHM}^2}/{8 {\ln} 2} },
\ee
where $\Delta_{XY}$ is the instrumental noise (in $\mu$K-radian) and
$\theta_\mathrm{FWHM}$ is the beam size (in radians). We consider a $1'$ beam,
$\Delta_{TT}=1\,\mu$K$'$, $\Delta_{EE}=\Delta_{BB}=1.4\,\mu$K$'$, and
$\Delta_{TE}=\Delta_{TB}=\Delta_{EB}=0$ and use measurements in the multipole range $\ell
= 2 - 3000$.

\subsection{Data covariance}

Because all CMB temperature and polarization anisotropies are lensed by the same realization of the lensing
potential, the lensed CMB power spectra data are correlated across multipoles. The ensuing covariance can be well
described by a simple analytical model which has been tested on simulations
\cite{BenoitLevy:2012va}. Recently, this model has been
extended to capture correlations of the lensed CMB power spectra with the lensing
potential $C^{\hat \phi\hat \phi}_\ell$ reconstructed by a quadratic estimator
\cite{Green:2016cjr, Peloton:2016kbw, Motloch:2016zsl}. Here we describe salient features
of this model, more in-depth discussion and graphical representation of the resulting
covariances can be found in the references.

In this model the correlation matrix is split into ``Gaussian part'' $\mathcal{G}$ that is diagonal
in multipole space and  $\mathcal{N}$ which describes non-Gaussian correlations between multipoles,
\be
\label{full_covariance}
	\Cov^{XY,WZ}_{\ell \ell'} = \mathcal{G}^{XY,WZ}_{\ell \ell'}
	+\mathcal{N}^{XY,WZ}_{\ell \ell'} .
\ee

The Gaussian part is modeled after the covariance of Gaussian random fields as
\be
\label{gaussian_covariance}
	\mathcal{G}^{XY, WZ}_{\ell \ell'} = \frac{\delta_{\ell \ell'}}{2\ell + 1}
	\left[C_{\mathrm{exp},\ell}^{XW}C_{\mathrm{exp},\ell}^{YZ} +
		C_{\mathrm{exp},\ell}^{XZ} C_{\mathrm{exp},\ell}^{YW}\right] ,
\ee
where the expectation value of the experimentally measured lensed CMB power spectra
$C^{XY}_\mathrm{exp}$ includes the noise power spectrum $N_\ell^{XY}$
\be
	C_{\mathrm{exp},\ell}^{XY} = C_\ell^{XY} + N_\ell^{XY}.
\ee

Even if we assume that the unlensed CMB fields $\tilde X$ and $\phi$ are Gaussian, the
lensed CMB fields $X$ are not. In our model, we take two non-Gaussian terms to compose the
full covariance,
\be
\label{nongauss_covariance}
	\mathcal{N}^{XY,WZ}_{\ell\ell'} =  \mathcal{N}^{(\phi)XY,WZ}_{\ell\ell'} +
	 \mathcal{N}^{(E)XY,WZ}_{\ell\ell'} .
\ee
The first term is
\be
\label{nongaussian_covariance}
	 \mathcal{N}^{(\phi)XY,WZ}_{\ell \ell'} = \sum_{L}
	\frac{\partial{C_\ell^{XY}}}{\partial C_{L}^{\PP}}\Cov^{\PP}_{LL}
	\frac{\partial{C_{\ell'}^{WZ}}}{\partial C_{L}^{\PP}}
\ee
and corresponds to correlations induced by the common set of gravitational lenses.
The power spectra derivatives are in practice calculated using a two point central difference scheme
from results obtained using CAMB\footnote{http://camb.info} \cite{Lewis:1999bs}. 

Sample variance of the unlensed $\tilde E \tilde E$ power spectrum and its coherent
propagation into the lensed power spectra through gravitational lensing produces similar but 
typically weaker effects. Following \cite{BenoitLevy:2012va}
we include this contribution only for
$\Cov^{XY,BB}_{\ell\ell'}$ with 
\be
		 \mathcal{N}^{(E)XY,BB}_{\ell, \ell'} = \sum_{L}
		\frac{\partial{C_\ell^{XY}}}{\partial C_{L}^{\tilde X \tilde Y}}\Cov^{\tilde X
		\tilde Y, \tilde E \tilde E}_{L,L}
		\frac{\partial{C_{\ell'}^{BB}}}{\partial C_{L}^{\tilde E \tilde E}}.
\ee
Other sample covariance effects from unlensed fields on $XY$ 
are negligible in comparison \cite{BenoitLevy:2012va}.   

\subsection{Simulated data and their analysis}

To simulate lensed CMB data we use the publicly available code 
Lenspix\footnote{https://github.com/cmbant/lenspix} 
\cite{Lewis:2005tp} {with unlensed CMB power spectra calculated by CAMB.} 
We modified the code to
lower its memory demands and speed up the calculation, see Appendix~\ref{sec:lenspix}
for details on these modifications. {Once the lensed CMB maps are generated, we add 
normally distributed\footnote{To avoid confusion between 
Gaussian covariance and Gaussian distributions, we use ``normal" for the latter.}}
instrumental noise and calculate power spectra to form a simulated data set
$\hat C^{XY}_\ell$. 

In most of the analyses in this paper we investigate the simulated CMB power spectra
using the Markov Chain Monte Carlo (MCMC) code CosmoMC\footnote{https://github.com/cmbant/CosmoMC}
\cite{Lewis:2002ah}, which for a given realization of the data samples the posterior
probability in the space of cosmological parameters. We assume uniform priors
in the cosmological parameters 
and use the likelihood described in the next section.  
In the MCMC runs we sample the posterior until the Gelman-Rubin statistic $R-1$
\cite{Gelman:1992zz} drops below 0.01.

To avoid biasing results, we calculate the unlensed fiducial spectra which enter Lenspix
simulations with the same precision settings which is later used in CosmoMC. We checked
that increasing precision with which the lensing operation in CosmoMC
is calculated (increasing \verb|accuracy_boost| in the lensing routine) does not
have any effect on the resulting parameter constraints. 

\subsection{Likelihood}

An accurate likelihood for CMB power spectra data $\hat C_\ell^{XY}$ has to capture both
lensing-induced covariance $\Cov^{XY,WZ}_{\ell,\ell'}$ and the non-normal distribution of 
the power spectra at low multipoles. Here we illustrate these effects with simulated data and then
describe our model for the likelihood.

Using 2000 Lenspix simulations, it is possible to illustrate that the lensed CMB data are
indeed correlated. Because this number of simulations is insufficient to show correlation
of individual power spectra multipoles, we look at correlation between band powers
\be
	P^{XY}_{\ell_1,\ell_2} = \frac{1}{\ell_2 - \ell_1} \sum_{\ell = \ell_1}^{\ell_2}
	\frac{\ell(\ell + 1) \Delta C^{XY}_{\ell}}{2\pi} ,
\ee
where
\be
\label{DeltaC}
	\Delta C^{XY}_\ell = \hat C^{XY}_{\ell} - C^{XY}_{\mathrm{exp}, \ell}
\ee
is the deviation of the experimentally measured power spectrum from its expectation value. As
an example, in Figure~\ref{fig:correlated_BB} we plot the distribution of two BB band powers
as determined from our simulations, together with theoretical curves showing {68\%
and 95\%} confidence intervals derived from our model for the covariance $\Cov^{XY,
WZ}_{\ell,\ell'}$.  We see that the data are indeed strongly correlated and that the model describes this correlation well.

\begin{figure}
\center
\includegraphics[width = 0.49 \textwidth]{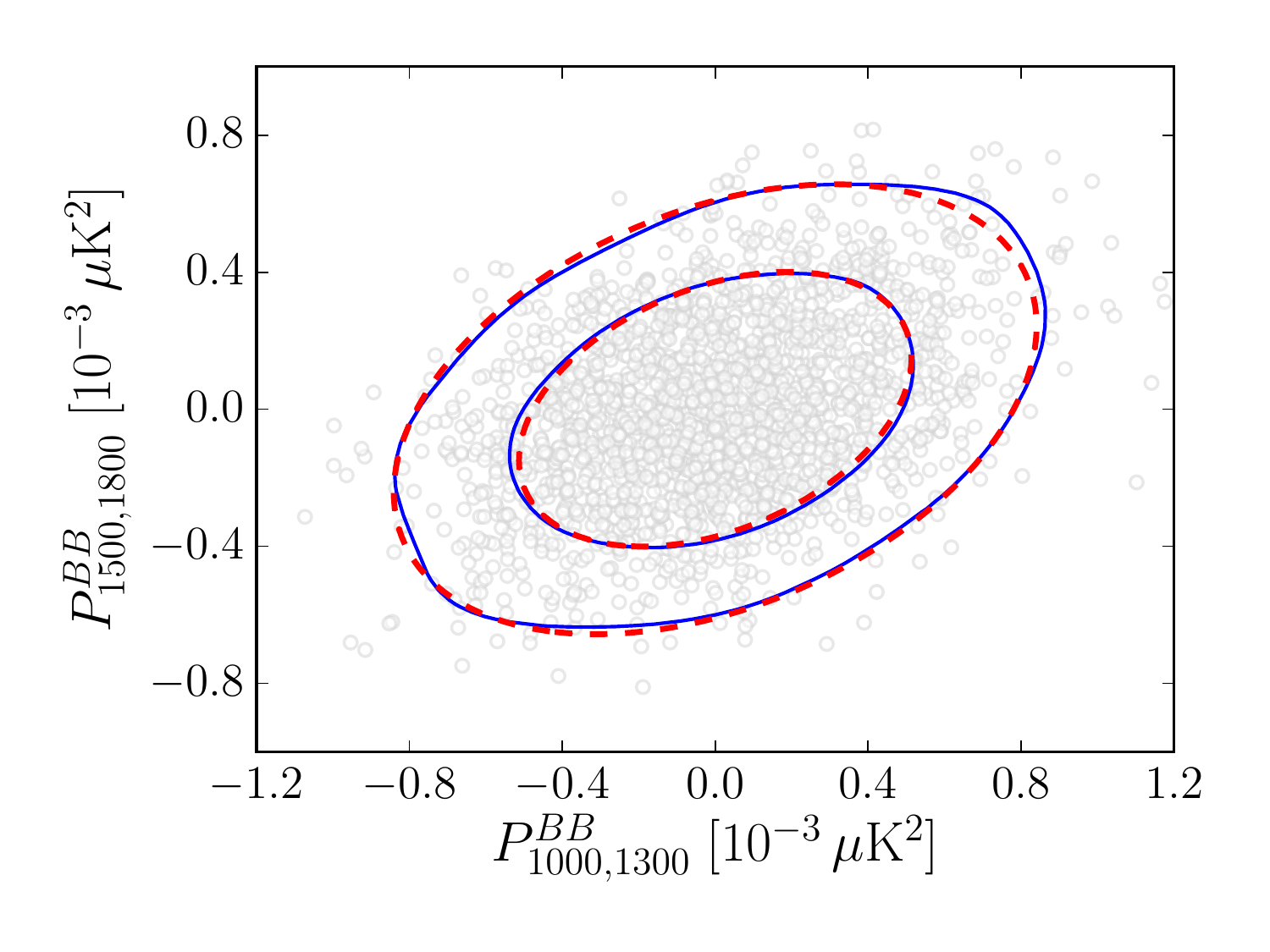}
\caption{Light gray circles show the correlated values of the binned power spectra
$P^{BB}_{1000,1300}$ and $P^{BB}_{1500,1800}$ as determined from 2000 Lenspix simulations.
Blue lines encompass regions of {68\% and 95\% confidence} determined from
these simulations. For comparison, dashed red lines show the same confidence intervals
based on our theoretical model for covariance.
}
\label{fig:correlated_BB}
\end{figure}

Fortunately for the likelihood construction, non-Gaussian covariances
$\mathcal{N}_{\ell\ell'}^{XY,WZ}$ of the low $\ell$ data are considerably smaller than the
corresponding Gaussian part of the covariance. Indeed, the largest correlation coefficient
\be
	R^{XY,WZ}_{\ell \ell'} =
	\frac
	{\mathcal{N}^{XY,WZ}_{\ell\ell'}}
	{\sqrt{\Cov^{XY,XY}_{\ell\ell}\Cov^{WZ,WZ}_{\ell'\ell'}}} 
\ee
with $\ell < 30$ is $8 \times 10^{-4}$. We shall therefore neglect
lensing-induced covariances in the likelihood for the large scale data.

As pointed out above, the low $\ell$ data are not normally distributed.
For example, in Fig.~\ref{fig:CBB_hist} we plot the distribution of 
$B$-mode power spectra  including noise for $\ell=2,30$ obtained in the same Lenspix simulations. In the
same plot we show a normal distribution centered on the expected mean of the data, with
standard deviation given by the Gaussian expectation \eqref{gaussian_covariance}
\be
	\sigma^{BB}_\ell = \sqrt{\frac{2 }{2 \ell + 1}}C^{BB}_{\mathrm{exp},\ell} .
\ee
It is clear that for $\ell=2$ the normal distribution is a poor description of the data. Instead, as
expected, $\chi^2_{2\ell+1}$ distribution scaled by $\sigma^{BB}_\ell$ fits the data well.
In the case of large scale $B$-modes this reflects the fact that they
get most of their power from $\tilde E, \phi$ modes at $\ell$ of several hundred.
Each coefficient in the spherical harmonic expansion of the $B$ map is then a combination of many random
fields and thus approximately normally distributed, which leads to $\chi^2$ distributed power
spectra.  For low $\ell$ $TT, TE$ and $EE$, the distributions just mirror the unlensed CMB fields, due
to negligible effects of lensing on these fields. Above $\ell \sim 30$ the normal distribution becomes a good
description {for} both the $\chi^2$ distribution and the data.

\begin{figure*}
\center
\includegraphics[width = 0.49 \textwidth]{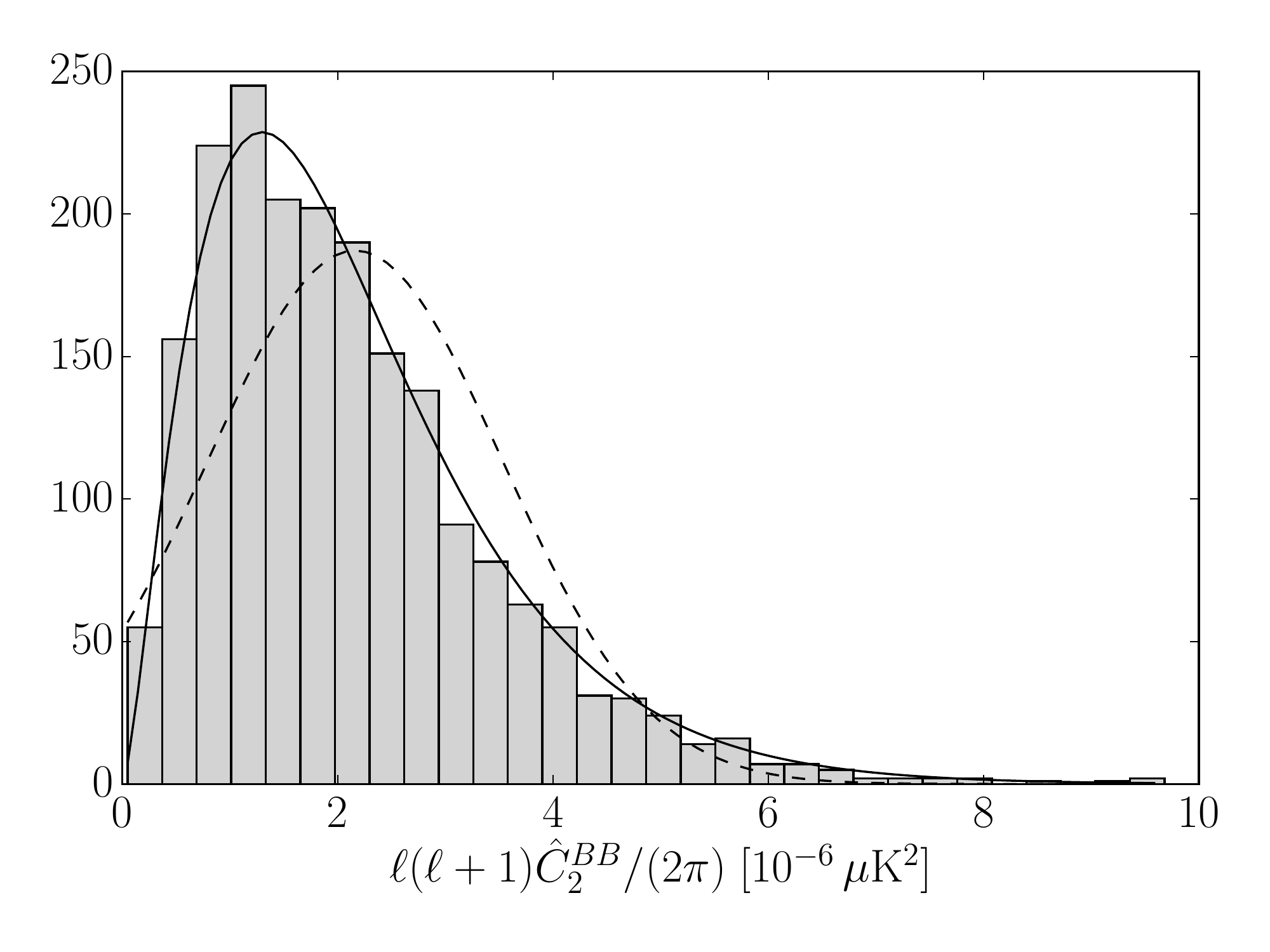}
\includegraphics[width = 0.49 \textwidth]{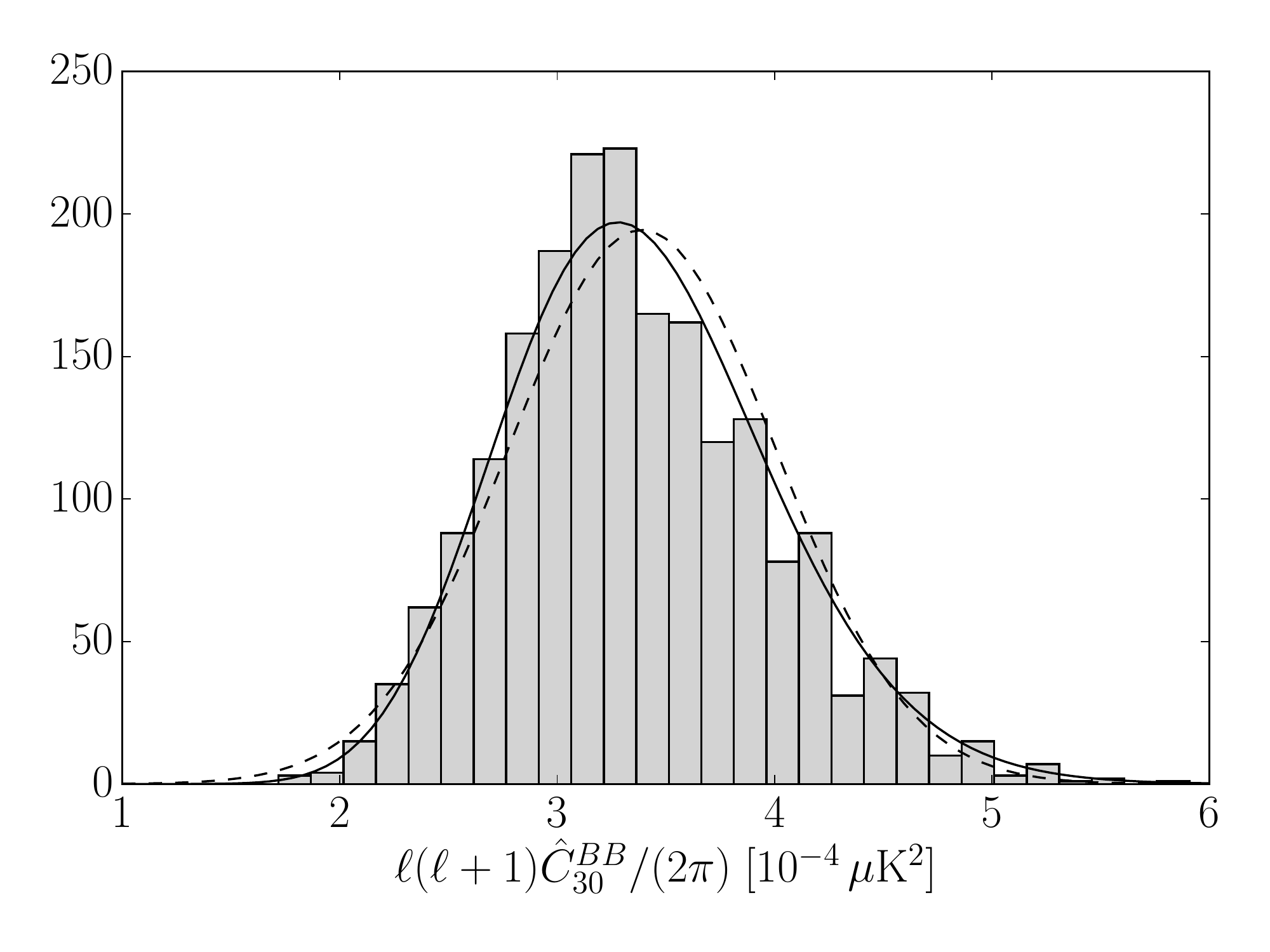}
\caption{Distribution of $\hat C^{BB}_{2}, \hat C^{BB}_{30}$ obtained from 2000 Lenspix simulations compared against a $\chi^2$ distribution with the Gaussian variance $(\sigma^{BB}_\ell)^2$
(solid) and a normal distribution with the same variance centered on the {expected mean
of the data} (dashed).  The normal distribution becomes a good approximation for $\ell \gtrsim 30$.}
\label{fig:CBB_hist}
\end{figure*}

Based on these considerations, our model for the likelihood treats $\hat C_\ell^{XY}$ from the largest scales ($\ell <
\ell_\mathrm{break}$) and from smaller scales  separately and
independently. The choice of the division point is
somewhat arbitrary; in this work we use $\ell_\mathrm{break} = 30$.

Below $\ell_\mathrm{break}$ we neglect lensing-induced covariance
$\mathcal{N}^{XY,WZ}_{\ell,\ell'}$ and assume there is no correlation between $B$ modes
and $T,E$ modes. Data with different multipoles $\ell, \ell'$ then decouple. The likelihood 
of measuring the data vector $\hat C_\ell^{XY}$ (including instrumental noise)
when the expected power spectra are $C^{XY}_{\mathrm{exp}, \ell}$ is
then a  product of inverse Wishart distributions  
\ba
\label{low_ell_likelihood}
	\mathcal{L}_{\ell<\ell_\mathrm{break}}& \propto & 
	{\prod}_{\ell = 2}^{\ell_\mathrm{break}-1}
	\left|C^{XY}_{\mathrm{exp},\ell}\right|^{-(2\ell + 1)/2} \\
	&& \times
	\exp\(
	-\frac{{2\ell + 1}}{2} \sum_{X,Y}
	\(C_{\mathrm{exp},\ell}^{XY}\)^{-1}
	{\hat{C}}^{XY}_\ell
	\)  . \nonumber
\ea
Here $| \cdot |$ is a determinant of $\cdot$ viewed as a matrix, here in the
 $X,Y$ space.

For $\ell \ge \ell_\mathrm{break}$ we neglect the non-normality of the distribution
of each multipole and
instead model the lensing-induced covariance between multipoles:
\ba
\label{high_ell_likelihood}
	\lefteqn{\mathcal{L}_{\ell \ge \ell_\mathrm{break}} \propto |\Cov|^{-{1/2}}
} \\
	&&	\exp\Bigg[ -\frac{1}{2}\sum_{
	\substack{
		\text{$ \ell,\ell' \ge \ell_\mathrm{break}$}
		\\
		\text{$XY, WZ$}
	}}
	\Delta C^{XY}_\ell
	\(\Cov^{XY,WZ}_{\ell, \ell'}\)^{-1}
	\Delta C^{WZ}_{\ell'}
	\Bigg]. \nonumber
\ea
In all analyses in this paper we neglect the dependence of the covariance matrix 
on the cosmological parameters by evaluating it at the fixed fiducial model of Tab.~\ref{tab:fiducial}. 

To assess the impact of  non-Gaussian covariance, we also investigate the likelihood
$\mathcal{L}_{\mathrm{g},\ell \ge \ell_\mathrm{break}}$ in which the non-Gaussian covariance $\Cov_{\ell,\ell'}$ is
replaced by the Gaussian covariance $\mathcal{G}_{\ell,\ell'}$.

By joining the large and small scale portions independently, we then form the total likelihood for the data 
\be
\label{lik}
	\ln \mathcal{L} = 
	\ln \mathcal{L}_{\ell < \ell_\mathrm{break}} +
	\ln \mathcal{L}_{\ell \ge \ell_\mathrm{break}} 
\ee
and
\be
\label{lik_g}
	\ln \mathcal{L}_\mathrm{g} = 
	\ln \mathcal{L}_{\ell < \ell_\mathrm{break}} +
	\ln \mathcal{L}_{\mathrm{g}, \ell \ge \ell_\mathrm{break}} ,
\ee
up to irrelevant additive constants.

By comparing analyses based on these two likelihoods we are able to gauge the impact of 
non-Gaussian covariance on parameter constraints. The lensing-induced
covariance is an additional source of correlated noise so constraints based on the likelihood with
Gaussian covariance are typically too optimistic.

\section{Cosmological parameter estimation}
\label{sec:cosmological_parameters}

In this section we investigate how neglecting lensing-induced covariance affects 
constraints on cosmological parameters. We focus here on the six
parameter flat $\Lambda$CDM cosmological model and two extensions where either
the  sum of the masses of neutrinos $\sum m_\nu$ ($\Lambda$CDM+$\sum m_\nu$) 
or 
the dark energy equation of state parameter $w$ ($\Lambda$CDM+$w$) 
is allowed to vary. 
Similar studies on the impact of non-Gaussian covariances have been previously
performed mainly using the Fisher approximation \cite{Smith:2006nk, BenoitLevy:2012va,
Motloch:2016zsl,Peloton:2016kbw}. We find that in at least one case ($\Lambda$CDM+$w$) the
Fisher approximation significantly underestimates the impact of lensing-induced covariance.

For $\Lambda$CDM we also illustrate how neglecting lensing-induced covariance leads to
{a significant increase in the fraction of realizations in which the fiducial model
parameters are excluded at 95\% confidence, which is potentially important for
concordance studies.

\subsection{$\Lambda$CDM}

Neglecting lensing-induced covariances for a typical simulated CMB dataset affects
constraints on $\Lambda$CDM parameters as shown in Fig.~\ref{fig:lcdm}. Shifts in the best
fit parameter values of the base parameters of Tab.~\ref{tab:fiducial} are typically not
very significant; the major effect of including lensing
covariances is a weakening of the best constrained directions between degenerate
parameters, most notably that between $\Omega_c h^2$ and $A_s$. We comment on the origin
of this effect in \S\ref{sec:explaining}.

\begin{figure}
\center
\includegraphics[width = 0.50 \textwidth]{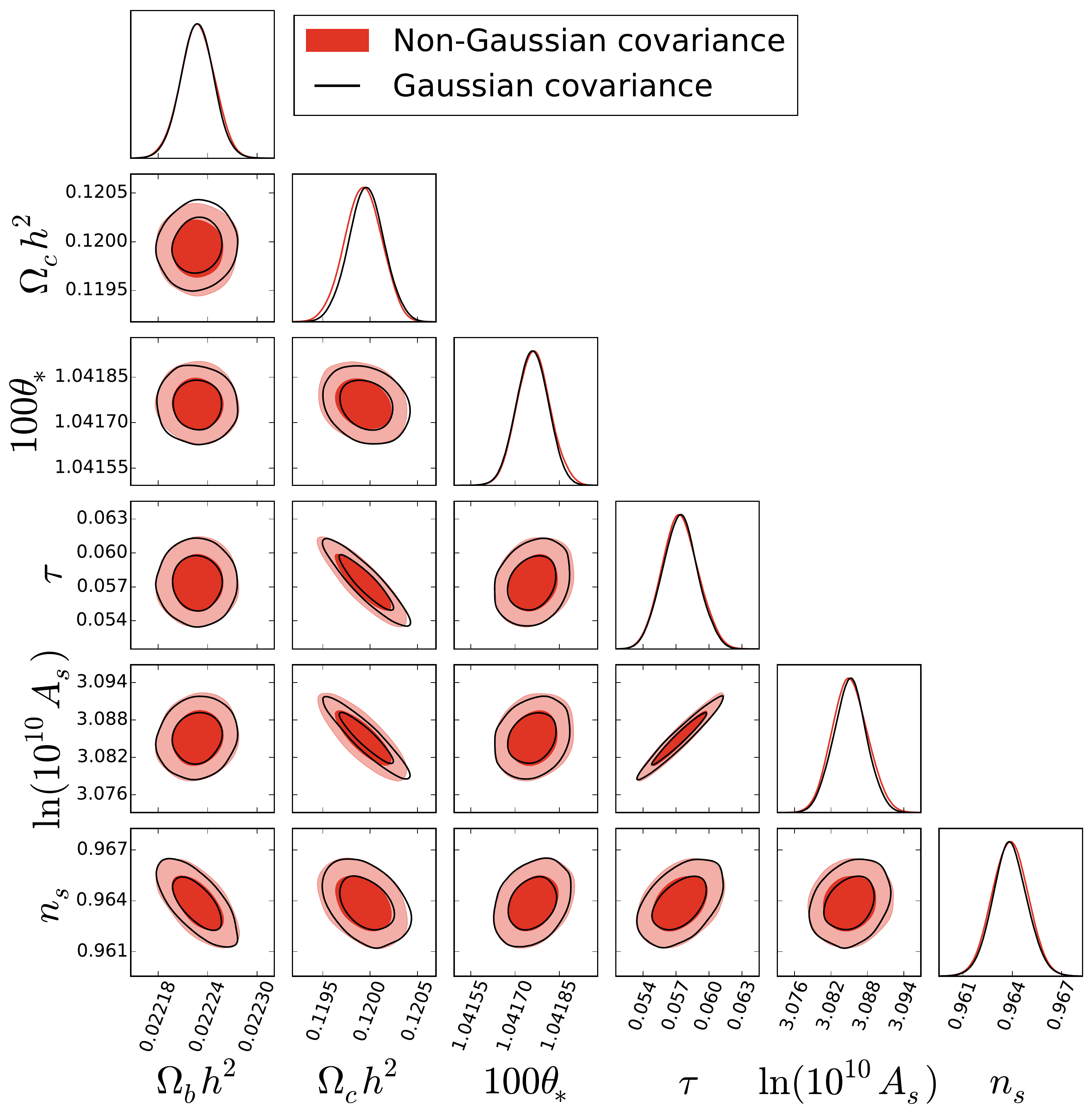}
\caption{Comparison of MCMC constraints on $\Lambda$CDM parameters with analysis based on
Gaussian ({black} curves) and non-Gaussian covariance ({red shaded}). Here and throughout contours
enclose regions of 68\% and 95\% confidence intervals unless otherwise specified. 
}
\label{fig:lcdm}
\end{figure}

Because of marginalization of other parameters, the two parameter posteriors   in Fig.~\ref{fig:lcdm}  
hide some of the effects of the non-Gaussian covariance. To uncover the maximal possible
effect on a single quantity, we construct the linear combination of the cosmological parameters
\be
\label{neutrino_combination}
	M = \sum_A \mathcal{K}_A \(\theta_A - \theta_A^\mathrm{fid}\)
\ee
{that maximizes the ratio of Gaussian to non-Gaussian errors.  {Here}
$\mathcal{K}_A=\{ 5.8,-13.4,18.4,-1.1,-2.6,3.1\}$ for the
parameter ordering $\{{{100}\theta_*},{\Omega_c h^2},{\Omega_b h^2},{n_s},{{\ln A_s}},{\tau}
\}$.
As we {discuss} in greater detail in \S\ref{sec:explaining}, $M$ can be interpreted as a
combination of cosmological parameters which mainly changes the lensing potential, especially at low $\ell$.

{Dashed} red lines in Fig.~\ref{fig:M_hist} show posterior probabilities
for $M$, as determined from a single MCMC run based on $\mathcal{L}$ (left) and
$\mathcal{L}_\mathrm{g}$ (right). The same simulated CMB
sky as in Fig.~\ref{fig:lcdm} {is} used, the maximum has been shifted to zero for future convenience,
and the $y$-axis units are arbitrary. The standard deviations of these two posteriors
are $\sigma^\mathrm{ng}_M = {2.1} \times 10^{-3}$ and $\sigma^\mathrm{g}_M = {1.0} \times
10^{-3}$, both within 3\% of the Fisher forecast prediction {displayed in
Fig.~\ref{fig:M_hist} by a solid blue line}.
The analysis based on a Gaussian likelihood $\mathcal{L}_\mathrm{g}$ therefore underestimates the errors 
of $M$ by over a factor of 2.

While the impact of this direction is hidden in the marginalized errors of the base parameters, it
reveals itself in an increased frequency of Type 1 errors: falsely rejecting the true model.
  This 
effect would be particularly problematic for concordance studies searching for
tensions between various cosmological datasets.  

\begin{figure*}
\center
\includegraphics[width = 0.49 \textwidth]{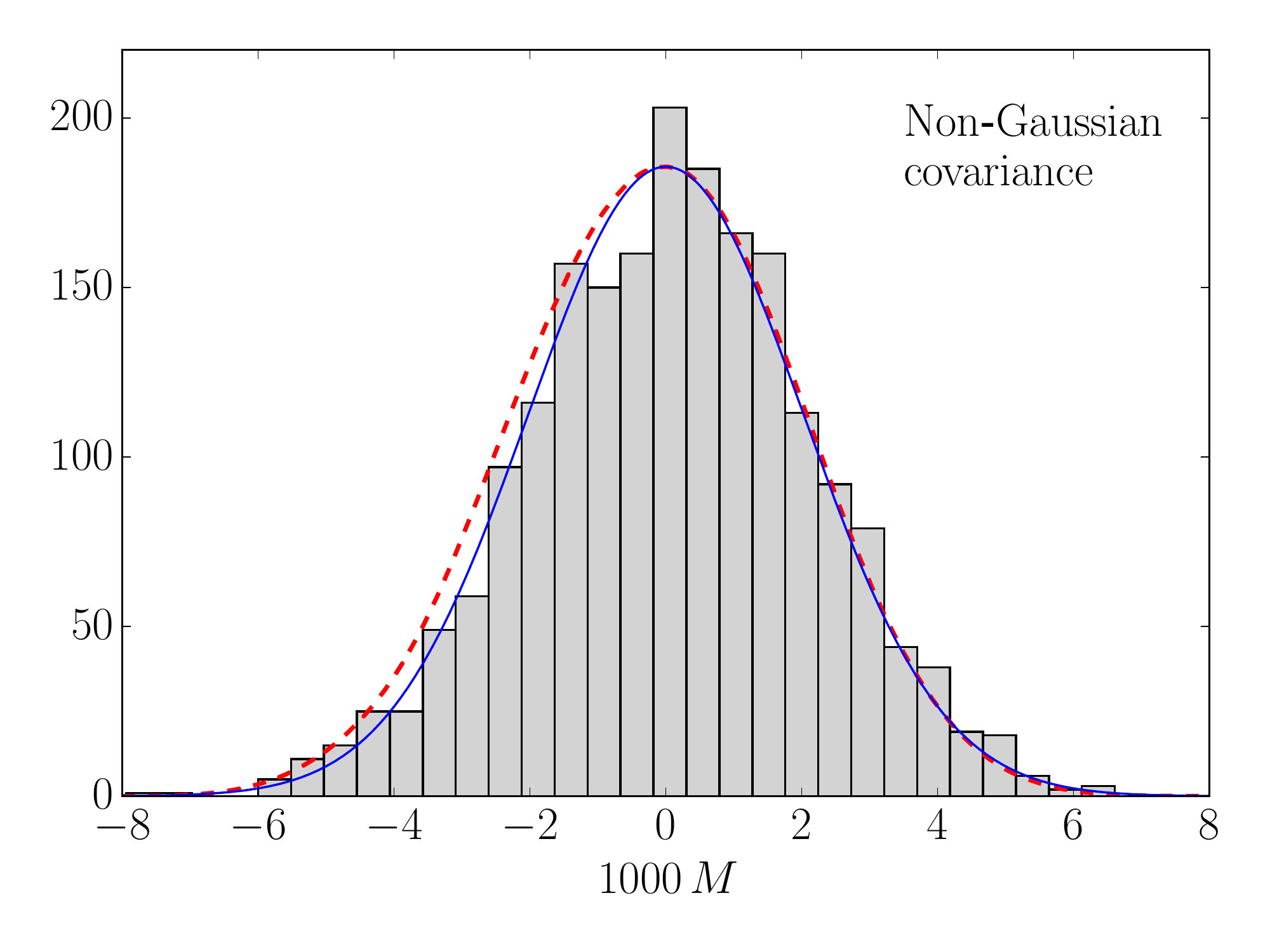}
\includegraphics[width = 0.49 \textwidth]{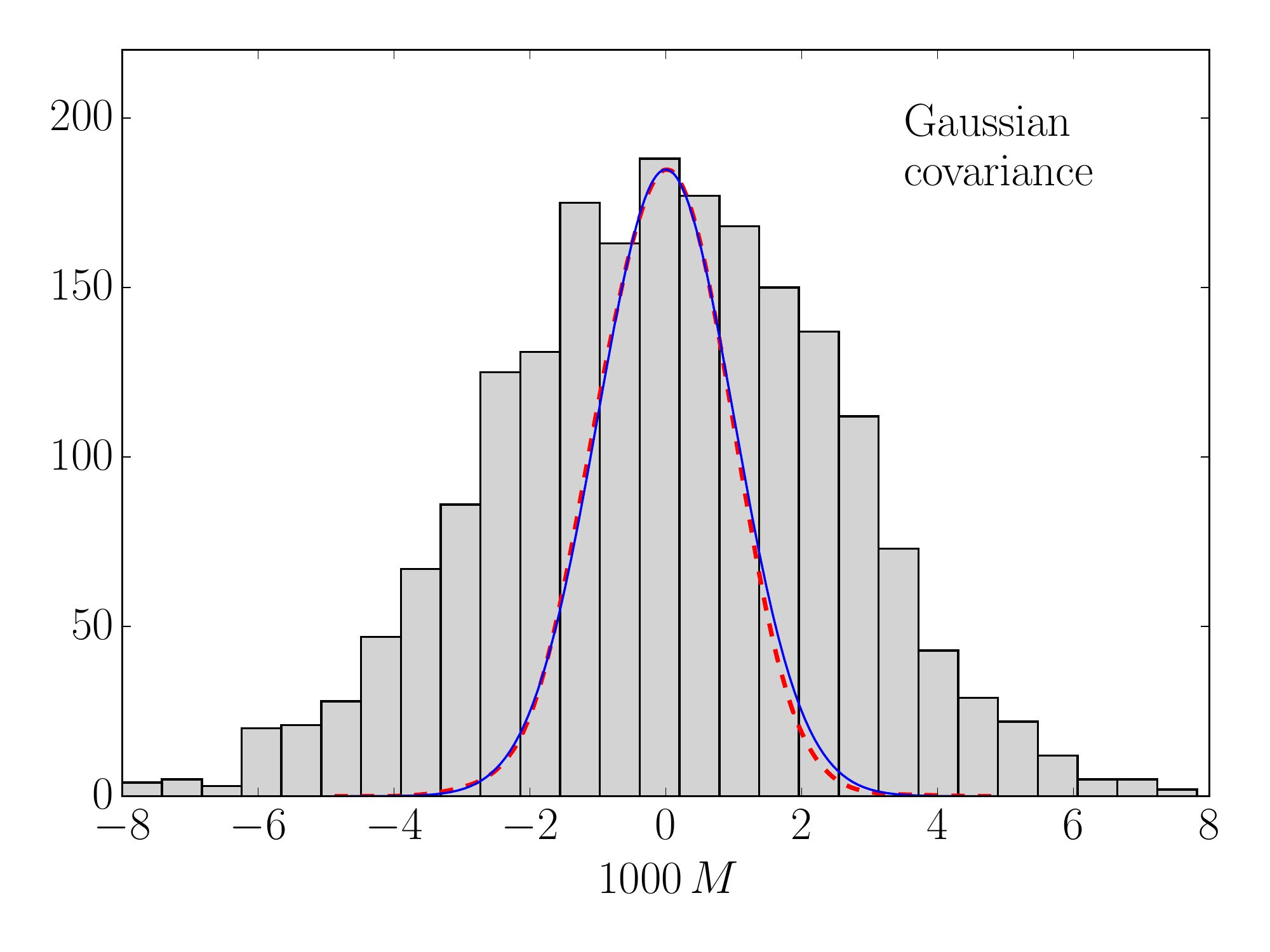}
\caption{Histograms showing best fit values of $M$ determined from 2000 lensed CMB skies
using an analytic approach (see text) for analysis based on non-Gaussian (left) and
Gaussian covariance (right). The dashed red curves show the 1D posterior probability for $M$,
as determined from a single CMB realization, shifted to zero for better comparison.
 Despite the Fisher forecast (solid blue curves) agreeing in both cases, the posterior probability does not reflect the much wider best-fit distribution in the Gaussian case.}
\label{fig:M_hist}
\end{figure*}

Since a full study of Type 1 errors in thousands of data realizations is computationally expensive, 
we illustrate this problem by analytically approximating the best fit values of the parameters,
including $M$, for each realization. 
We assume the data are sufficiently close to the fiducial model that it is possible to
approximate the $C_\ell^{XY}$ as linear in the parameter deviations  $\theta_A-\theta_A^{\rm fid}$.
Neglecting for the moment complications arising from presence of the low $\ell$
data by assuming all the data are distributed according to a multivariate normal
distribution with covariance \eqref{full_covariance}, we obtain the maximum likelihood or best fit estimate for a cosmological parameter
$\theta^\mathrm{bf}_A$ as\footnote{Numerical maximization of the likelihood in ten simulations provides an average shift in $M$ below $0.08\,
\sigma_{M,\mathrm{bf}}$  with respect to the analytic formula  \eqref{analytic_bestfit} for either covariance. The analytic treatment is thus sufficiently accurate for our purposes.}
\begin{eqnarray}
\label{analytic_bestfit}
	\Delta \theta_A &\equiv&
	 \theta_A^\mathrm{bf} - \theta_A^\mathrm{fid} \\
	 \nonumber  
	 &=&
	\sum_{
			\substack{
				\text{$ B,i,j$}
			}}
	F^{-1}_{AB} \frac{\partial D_i}{\partial \theta_B}
	\( \Cov^{-1}\)_{ij}\(\hat D_j - D_j^{ \mathrm{fid}}\) .
\end{eqnarray}
Here the power spectrum data are indexed as $D_i =  C^{XY}_\ell$ with $i$ running
over all $XY,\ell$ elements.
The Fisher information matrix,
\be
\label{fisher}
	F_{AB} =
	\sum_{
			\substack{ij}}
	\frac{\partial D_i}{\partial \theta_A}
	\(\Cov\)^{-1}_{ij}
	\frac{\partial D_j}{\partial \theta_B} ,
\ee
and parameter derivatives are evaluated around the fiducial model.

 With the best fit values of
cosmological parameters we can directly calculate the best fit values of $M$ for both the non-Gaussian and Gaussian analysis.  The two differ only in the choice of covariance matrix in Eqs.\ \eqref{analytic_bestfit} and
\eqref{fisher}.

Best fit values of $M$ determined from 2000 simulated CMB skies are shown in histograms in
Fig.~\ref{fig:M_hist} for the non-Gaussian (left) and Gaussian (right) covariance analyses.
In the non-Gaussian case, this distribution has a standard deviation
$\sigma^\mathrm{ng}_{M,\mathrm{bf}} = 2.1 \times 10^{-3}$ which is in excellent agreement with the prediction from the posterior $\sigma^\mathrm{ng}_M$ determined from the MCMC analysis of single realization as well as the Fisher approximation. On the other hand, in the
Gaussian analysis the best fit values of $M$ scatter with standard deviation
$\sigma^\mathrm{g}_{M,\mathrm{bf}} = 2.5 \times 10^{-3}$, which is 2.5 times
the width of $\sigma^\mathrm{g}_M$ despite the latter agreeing with its Fisher approximation. This mismatch can lead to Type 1 errors in cases
where the best fit $M$ fluctuates away from the fiducial value zero.

Notice that the best fit distribution is wider in the Gaussian than non-Gaussian 
case by a factor of $\sim 1.2$ which further exacerbates the probability of Type 1 errors. This
is not surprising, 
Eq.~\eqref{analytic_bestfit} {is a minimum variance estimator only if the assumed model of the covariances is correct which it is not in the Gaussian case.}

\begin{figure}
\center
\includegraphics[width = 0.49 \textwidth]{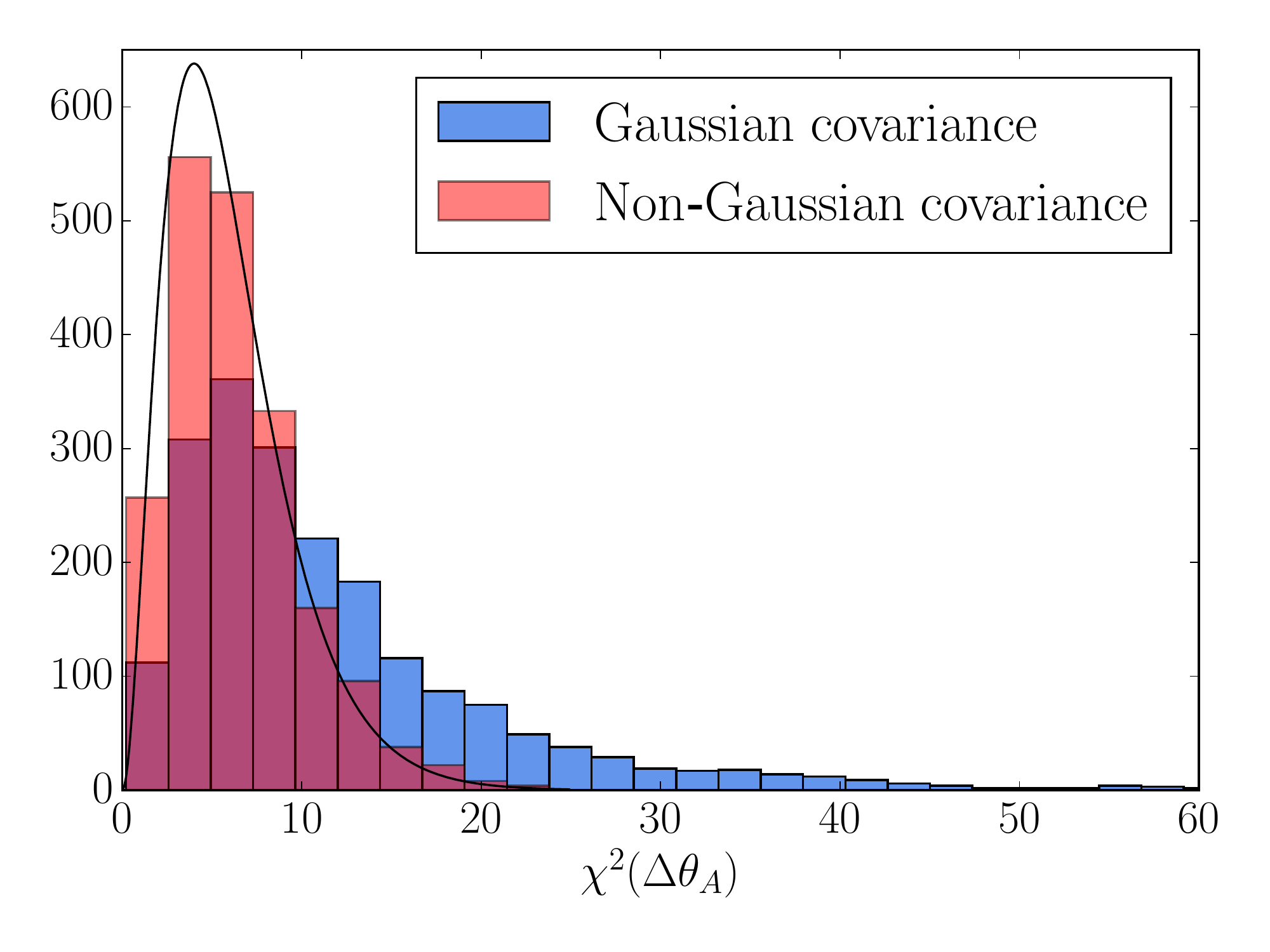}
\caption{{Histogram of $\chi^2(\Delta \theta_A)$ for the parameter deviations of the best
fit from the true model  \eqref{lik_fid}, as determined from
our simulations with Gaussian (blue) and non-Gaussian (red) covariance. For comparison, 
the solid line is proportional to probability density function of $\chi^2_6$.}  The long tail in the
Gaussian case leads to anomalously frequent Type 1 errors where the true model is rejected at high
confidence (see text).} 
\label{fig:lcdm_chi2_g_ng}
\end{figure}

To quantify the probability of Type 1 errors considering all parameters that specify the
$\Lambda$CDM model,  we can also compute $\chi^2$ between the best fit and the true
fiducial model assuming the errors from the posterior
\be
\label{lik_fid}
	\chi^2(\Delta \theta_A) 
	= \sum_{AB} \Delta \theta_A {\rm Cov}_{AB}^{-1} \Delta \theta_B 
	\approx \sum_{AB} \Delta \theta_A F_{AB} \Delta \theta_B .
	\ee
In order to estimate $\chi^2$ for each of the 2000  lensed CMB simulations, we again use the Fisher matrix
as an approximation to the inverse covariance.\footnote{Using the ten simulations in which
we find actual best fit $\Lambda$CDM parameters {and use actual posterior covariance
for the parameters}, we can estimate what is the error from
using in \eqref{lik_fid} the analytic estimator
{for $\Delta \theta_A$ and Fisher matrix}. The average change in
$\chi^2(\theta_A)$ we observe is {0.7}, sufficiently smaller than the width of the final
distributions. {Most of this difference comes from the analytic estimator
\eqref{analytic_bestfit}.}
}
The variable $\chi^2(\Delta \theta_A)$ should be
$\chi^2_6$ distributed, where 6 is the number of cosmological parameters in $\Lambda$CDM. 

Histograms of $\chi^2(\Delta \theta_A)$ for the Gaussian and non-Gaussian analysis 
are compared in Fig.~\ref{fig:lcdm_chi2_g_ng} to the theoretical expectation.
It is clear that {in} the Gaussian analysis, the misestimate of the parameter covariance
as well as the suboptimal estimate of the best fit causes a strong 
disagreement with the expected $\chi^2_6$ distribution.

For example, when the analysis is based on the Gaussian covariance, more than 30\% of the
simulations show $\chi^2(\Delta \theta_A) > 12.59$; for $\chi^2_6$ this value is exceeded
only in 5\% of the cases. As pointed out above, this can be potentially dangerous
for concordance studies. The non-Gaussian covariance leads to much better agreeement
($\sim$~6.9\% of simulations have $\chi^2(\Delta \theta_A) > 12.59$) and moreover  there is no long
tail to very high $\chi^2(\theta_A)$.

{
In the Appendix~\ref{sec:no_low_ell} we comment on small changes to some of the conclusions of
this section for an experimental configuration which observes only part of the sky and
does not measure data at multipoles $\ell < 30$.
}

\subsection{$\Lambda$CDM + $\sum m_\nu$}
\label{sec:lcdm+onu}

In this section we release the neutrino mass from its fiducial value and investigate
a seven-parameter extension of $\Lambda$CDM.  {In this case, the two parameter contour
plots, shown in Fig.~\ref{fig:lcdm+onu_contour}, show much smaller effects of non-Gaussian
covariance than in $\Lambda$CDM, though the impact is visible in the lower limit for
 $\sum m_\nu$. Ref.~\cite{Peloton:2016kbw} also found only small effects for this case.

\begin{figure}
\center
\includegraphics[width = 0.50 \textwidth]{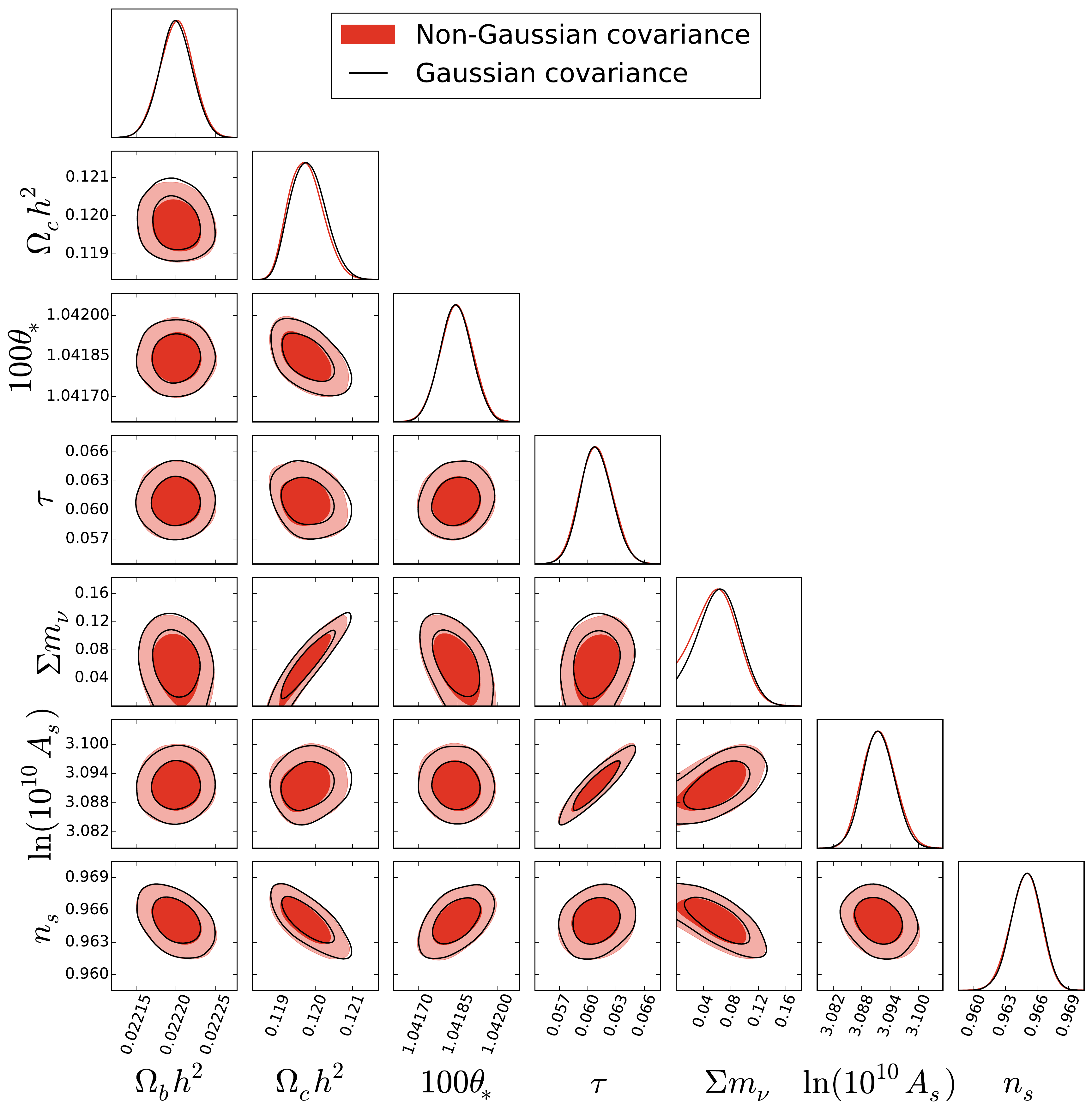}
\caption{Comparison of MCMC constraints on $\Lambda$CDM+$\sum m_\nu$ parameters with analysis based on
Gaussian (black curves) and non-Gaussian covariance (red shaded). 
}
\label{fig:lcdm+onu_contour}
\end{figure}

However, this does not mean non-Gaussian covariance can be neglected in this case, only
that its effects are hidden by marginalizations. As for $\Lambda$CDM, we can form a
combination of the cosmological parameters {which is} predicted by the Fisher forecast to show the largest effect of
non-Gaussian covariance,
\be
\label{neutrino_combination}
	M^{\nu} = \sum_A \mathcal{K}^{\nu}_A \(\theta_A - \theta_A^\mathrm{fid}\)
\ee
with $\mathcal{K}^{\nu}_A =
\{ 5.7,
 -13.3,
 18.4,
 -1.1,
 -2.6,
 3.1,
 0.29\, \mathrm{eV}^{-1}\}$, for the ordering
 $\{ {{100}\theta_*},
{\Omega_c h^2},
{\Omega_b h^2},
{n_s},
{{\ln A_s}},
{\tau},
{\sum m_\nu}\}$.
 In Fig.~\ref{fig:lcdm+onu} we can see 1D posterior probabilities
for $M^\nu$ from MCMC analyses based on Gaussian and non-Gaussian covariance. 
{ It is clear $M^\nu$ constraints are nearly as strongly affected by the lensing-induced
covariance as $M$ for $\Lambda$CDM, with standard deviation degrading from $1.1 \times 10^{-3}$ to $2.2 \times
10^{-3}$, even though this effect does not show up in any pair of base parameters. Likewise,
the Gaussian analysis is prone to Type 1 errors as in the $\Lambda$CDM case.
}

The standard deviations of $M^\nu$ quoted in the previous paragraph differ somewhat
from their Fisher forecasts: $1.0 \times 10^{-3}$ and $2.1 \times 10^{-3}$ for the
Gaussian and non-Gaussian cases respectively. This occurs in part because the
posterior is non-normal due to  the presence of the physicality prior $\sum m_\nu > 0$
which hides some of the non-Gaussian covariance effects (cf.
Fig.~\ref{fig:lcdm+onu_contour}).   In cases where neutrino mass is detected with high
significance and effects of the prior boundary is smaller, $M^\nu$ constraints are in good
agreement with the Fisher prediction.

\begin{figure}
\center
\includegraphics[width = 0.49 \textwidth]{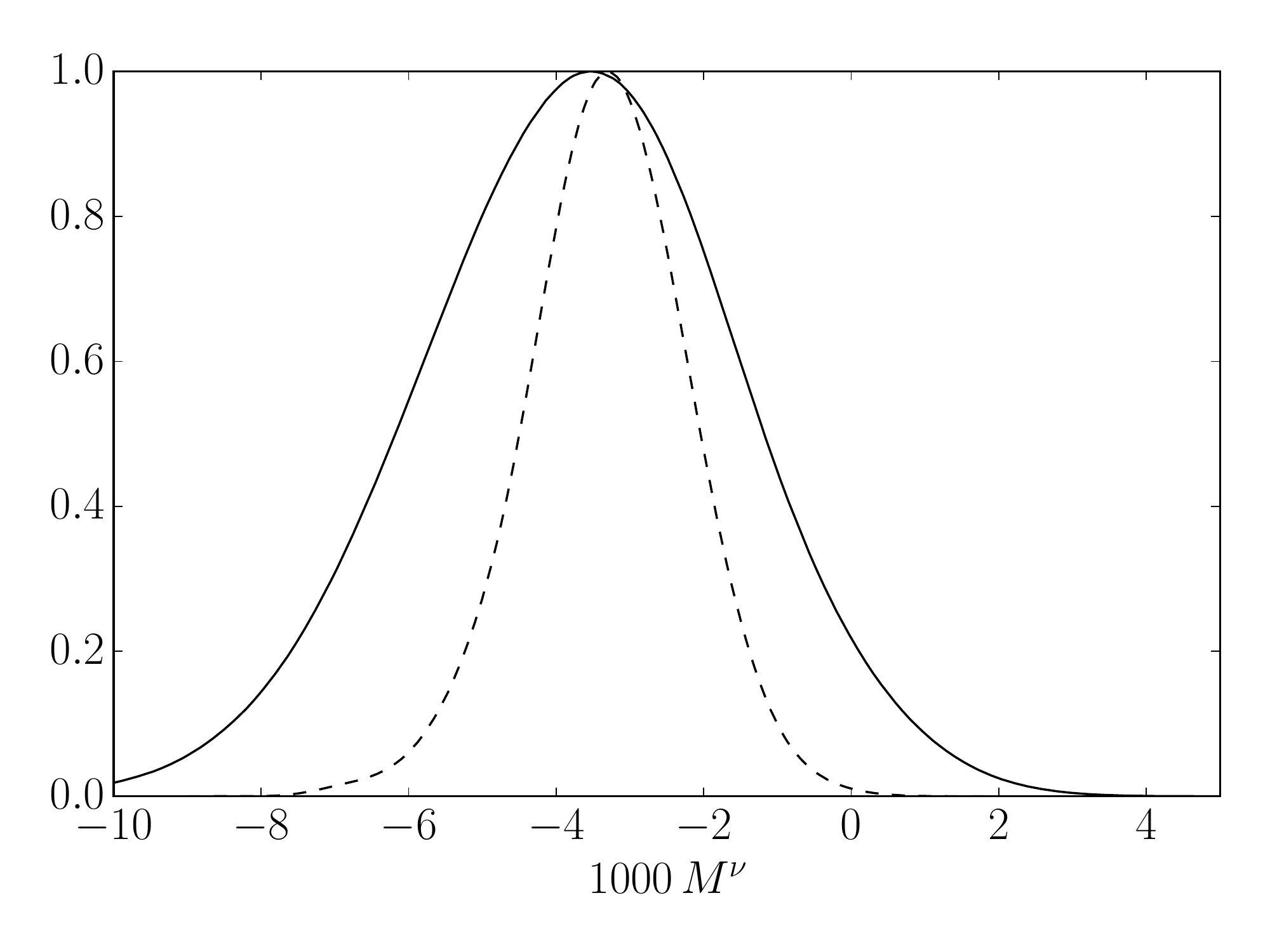}
\caption{Effect of the non-Gaussian covariance on constraints on a parameter combination
$M^\nu$ \eqref{neutrino_combination} within $\Lambda$CDM+$\sum m_\nu$; the combination was
chosen to maximize this effect. 
Solid line shows MCMC constraints with non-Gaussian
covariance, dashed line with the Gaussian covariance.}
\label{fig:lcdm+onu}
\end{figure}

\subsection{$\Lambda$CDM + $w$}

In the model where we allow the dark energy parameter of state to vary, the effect of
non-Gaussian covariance is more pronounced and clearly visible already on posterior
probability distribution for $w$, see Fig.~\ref{fig:lcdm+w}. The two analyses, based on
$\mathcal{L}$ and $\mathcal{L}_\mathrm{g}$, strongly disagree in the low $w$ tail; by neglecting
the covariance induced by the gravitational lensing one would wrongly rule out low values
of $w$. {For example, for this particular realization lower 95\% confidence limits
(two sided) for the non-Gaussian and Gaussian likelihoods are $-1.55$ and $-1.37$ respectively.} 
In \S\ref{sec:explaining} we look deeper into this behavior.

The impact of non-Gaussian covariance depends to some extent on best fit value
of $w$; simulations with lower best fit value of $w$ typically show larger effects of
non-Gaussian covariance. To illustrate this, in Fig.~\ref{fig:lcdm+w_scatter} we show
marginalized constraints on $w$ for six different simulations, both with Gaussian and
non-Gaussian likelihood. 

In the Fisher approximation, $\Lambda$CDM+$w$ was investigated for essentially the
same experimental configuration in \cite{Smith:2006nk}. It was found that non-Gaussian
covariances should increase errors on $w$ by only about 24\%, which is considerably less
than what we uncovered in the full MCMC analysis. In this case, the local approximation thus
gives misleading results, due to the non-normal posterior which the Fisher approximation
can not faithfully capture.
The origin of this behavior can be traced back to how dark energy affects lensing.
As $w$ decreases, its effect on the lensing potential quickly diminishes,
as dark energy ceases to be important at redshifts where the lensing kernel peaks. 
Because in this case the parameter combination
constrained by lensing is significantly changing throughout the allowed parameter
posterior, the Fisher analysis fails to capture the full significance of the non-Gaussian
covariance.

\begin{figure}
\center
\includegraphics[width = 0.50 \textwidth]{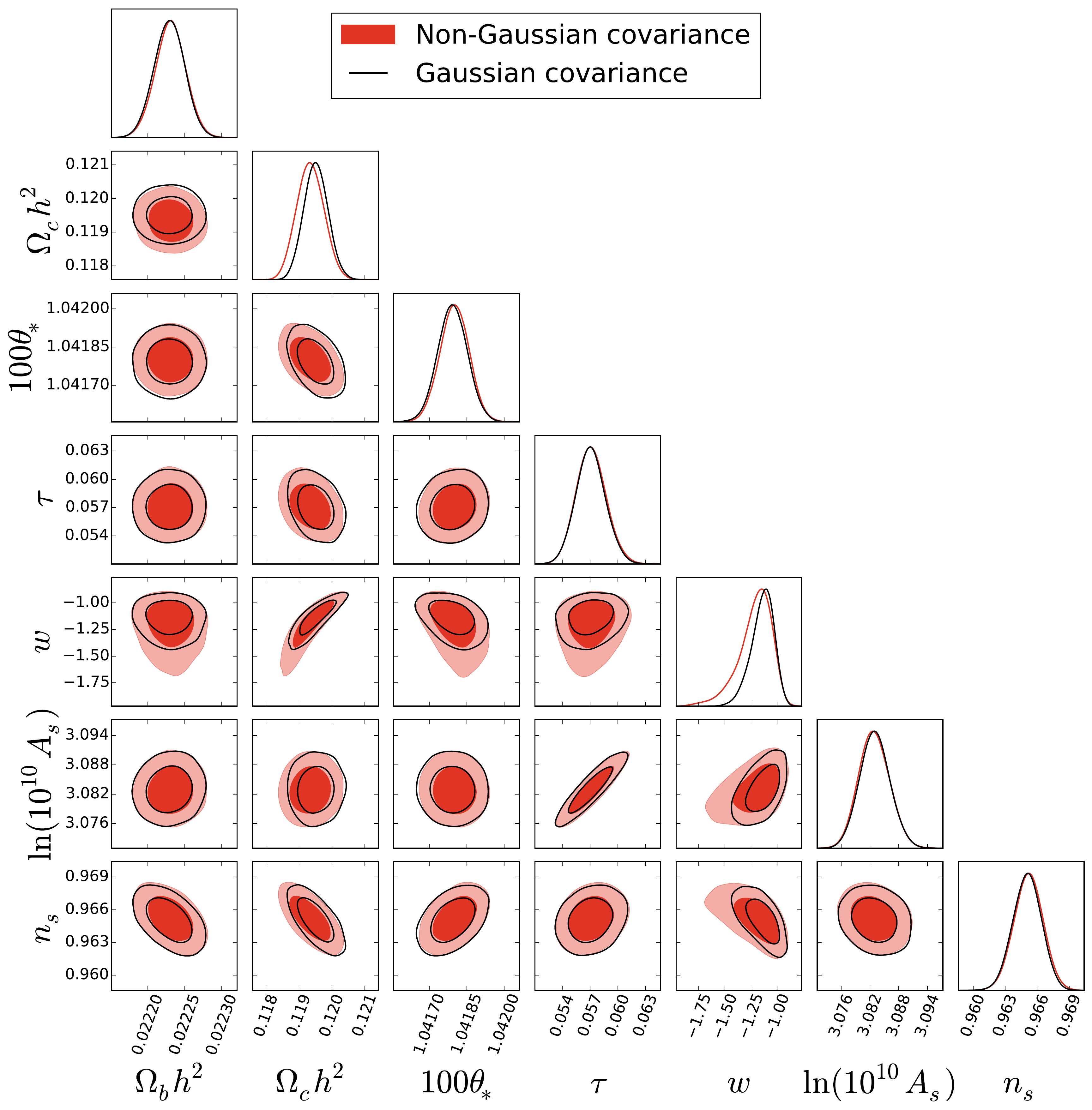}
\caption{Comparison of MCMC constraints on $\Lambda$CDM+$w$ parameters with analysis based on
Gaussian ({black} curves) and non-Gaussian covariance (red shaded).  The impact of
non-Gaussian covariance is clearly apparent in constraints involving $w$. 
}
\label{fig:lcdm+w}
\end{figure}

\begin{figure}
\center
\includegraphics[width = 0.49 \textwidth]{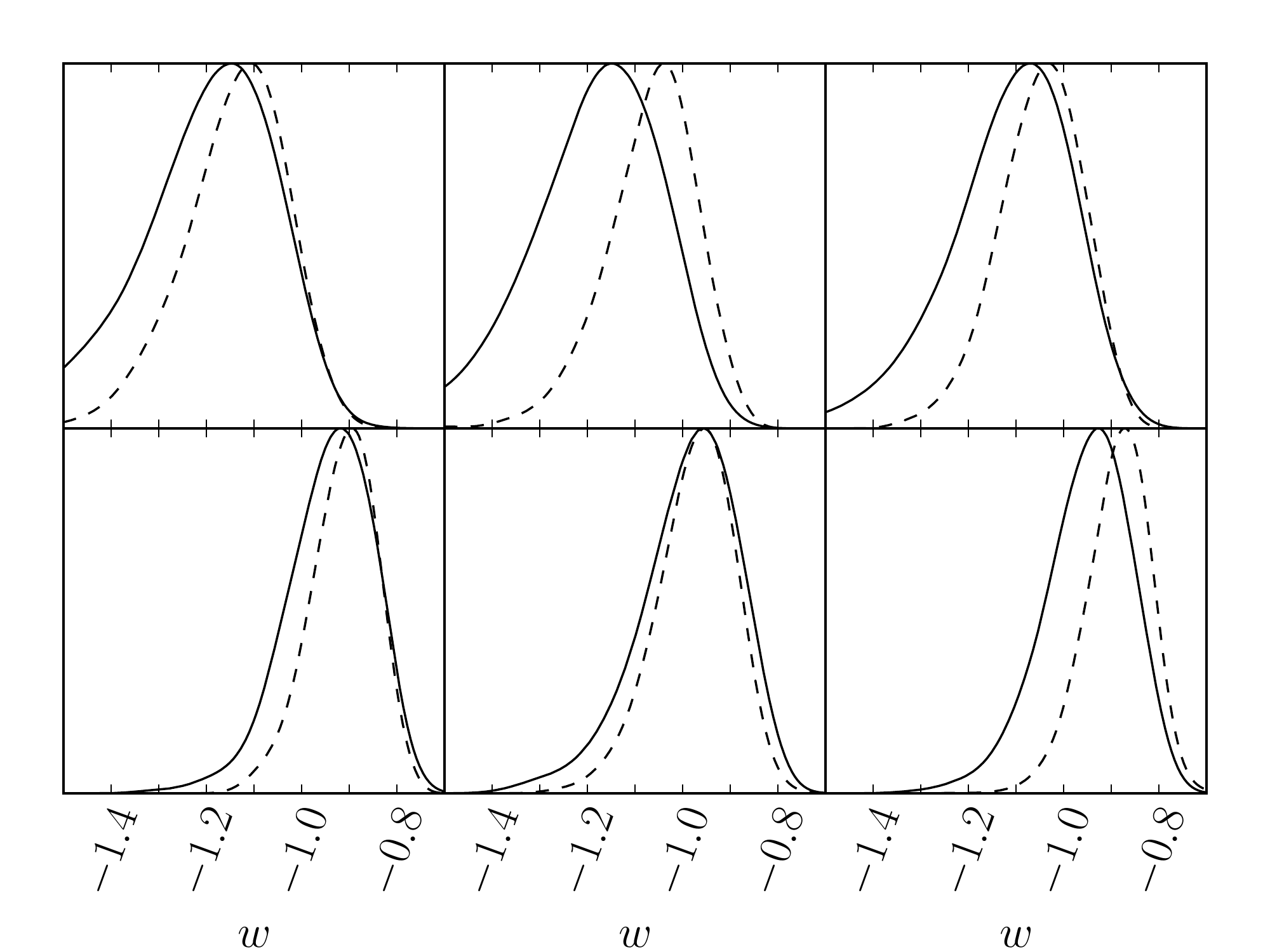}
\caption{Comparison of MCMC constraints on $w$ for our experimental setup;
each panel represents a different $\Lambda$CDM simulation of the CMB sky with
that of Fig.~\ref{fig:lcdm+w} in the top left corner. The Gaussian analysis
(dashed) increasingly deviates from the non-Gaussian analysis (solid) as the maximum likelihood
value of $w$ decreases.}
\label{fig:lcdm+w_scatter}
\end{figure}

The maximally impacted linear combination of base parameters also shows an enhanced
non-Gaussian effect compared with $\Lambda$CDM and $\Lambda$CDM+$\sum m_\nu$.
 It reads
\be
\label{w_combination}
	M^{w} = \sum_A \mathcal{K}^{w}_A \(\theta_A - \theta_A^\mathrm{fid}\) ,
\ee
where $\mathcal{K}^w_A=
\{ 5.7,
 -12.8,
 18.0,
 -1.0,
 -2.5,
 2.9,
 0.087\}$
 for the  ordering
 $\{{{100}\theta_*},
{\Omega_c h^2},
{\Omega_b h^2},
{n_s},
{{\ln A_s}},
{\tau},
{w}\} $.
Posterior probabilities for $M^w$ from MCMC analyses based on Gaussian and
non-Gaussian covariance are shown in Fig.~\ref{fig:M_lcdm+w}. In this case, the standard
deviations 
$\sigma^\mathrm{ng}_{M^w,\mathrm{Fish}} = 2.3 \times 10^{-3}$ 
and 
$\sigma^\mathrm{g}_{M^w,\mathrm{Fish}} = 1.0 \times 10^{-3}$
derived using a Fisher approximation show that the relative impact of non-Gaussian
covariance is in reasonable agreement with the MCMC {results}
$\sigma^\mathrm{ng}_{M^w} = 7.6 \times 10^{-3}$ 
and 
$\sigma^\mathrm{g}_{M^w} = 3.4 \times 10^{-3}$, but the overall scale is still strongly
underestimated by the Fisher analysis as  {is the extent of the lower tail}.  These mismatches
are expected for the same reason that they appear in $w$ alone, namely due to parameter
nonlinearity within the allowed region.  Likewise, $M^w$ defined by \eqref{w_combination}
only captures the parameter combination which is the most affected by the lensing-induced
covariance locally at the fiducial model parameters, not necessarily globally (see
\S\ref{sec:explaining}).
}

\begin{figure}
\center
\includegraphics[width = 0.49 \textwidth]{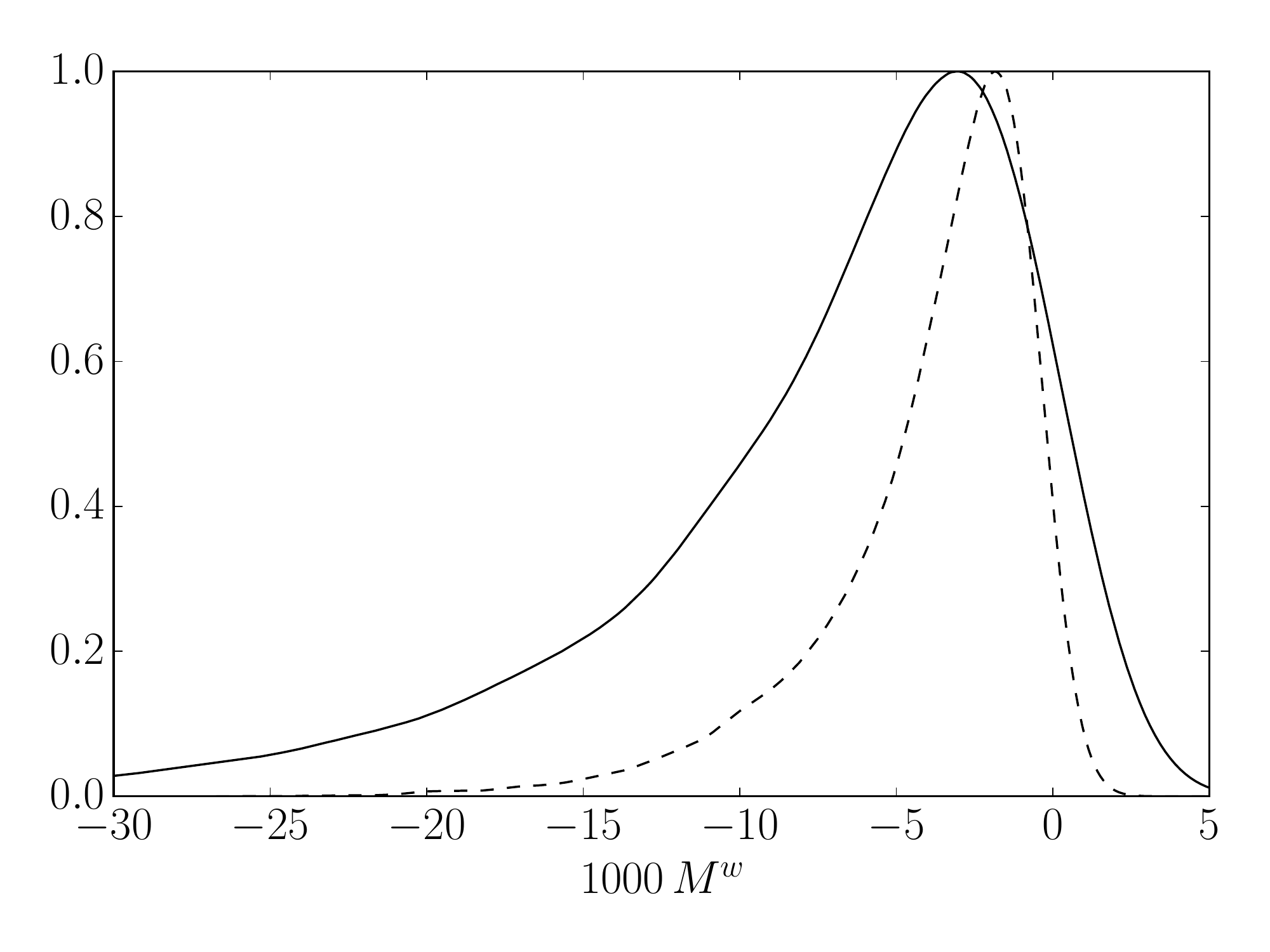}
\caption{Effect of the non-Gaussian covariance on constraints on a parameter combination
$M^w$ \eqref{w_combination} within $\Lambda$CDM+$w$; the combination was
chosen to maximize this effect locally around the fiducial model. 
{Solid line shows MCMC constraints with non-Gaussian
covariance, dashed line with the Gaussian covariance.}}
\label{fig:M_lcdm+w}
\end{figure}

\section{Lens principal components}
\label{sec:lens_pc}

The temperature and polarization power spectra contain more information about the lensing
potential than just its amplitude; this information can be faithfully captured in terms of
a few principal components (PCs) \cite{Motloch:2016zsl, Smith:2006nk}. This model-independent
lensing information can be utilized in a variety of ways. By isolating the lensing information,
we can more directly diagnose when lensing-induced
covariance is important for parameter constraints. Lensing PCs  also enable us to
check the internal consistency of the data with a given model, by comparing the constraints on parameters derived from the lensing potential
with those from the unlensed power spectra. Comparing
 $C^{\phi\phi}_\ell$ constraints from the power
spectra with those from the reconstructed lensing potential and external measurements
provides yet another powerful consistency check.  Finally, PCs enable model building beyond the currently considered model classes should these consistency tests fail.

In this section we conduct an MCMC study of the lensing PCs $\Theta^{(i)}$ as defined
by the Fisher approximation \cite{Motloch:2016zsl, Smith:2006nk} together
with variables $\tilde \theta_A$ that parametrize the unlensed power spectrum.  Since
the PCs are defined under the Fisher approximation, we first verify that in the MCMC analysis
the PCs remain unbiased and weakly correlated --
both mutually and with $\tilde \theta_A$. Then we point out that almost the whole effect of
the non-Gaussian covariance is manifested in the first two PCs $\Theta^{(1)}, \Theta^{(2)}$ and use this
knowledge to explain effects of non-Gaussian covariances on parameter constraints seen in
the previous section.

We end this section with a discussion of possible consistency checks using these principal
components and a suggestion of how 
to use them to compress most of the information contained in the lensed CMB power spectra into
a simple normal likelihood, which can be used to quickly determine approximate constraints
on a wider class of models than explicitly analyzed here.

\begin{figure}
\center
\includegraphics[width = 0.49 \textwidth]{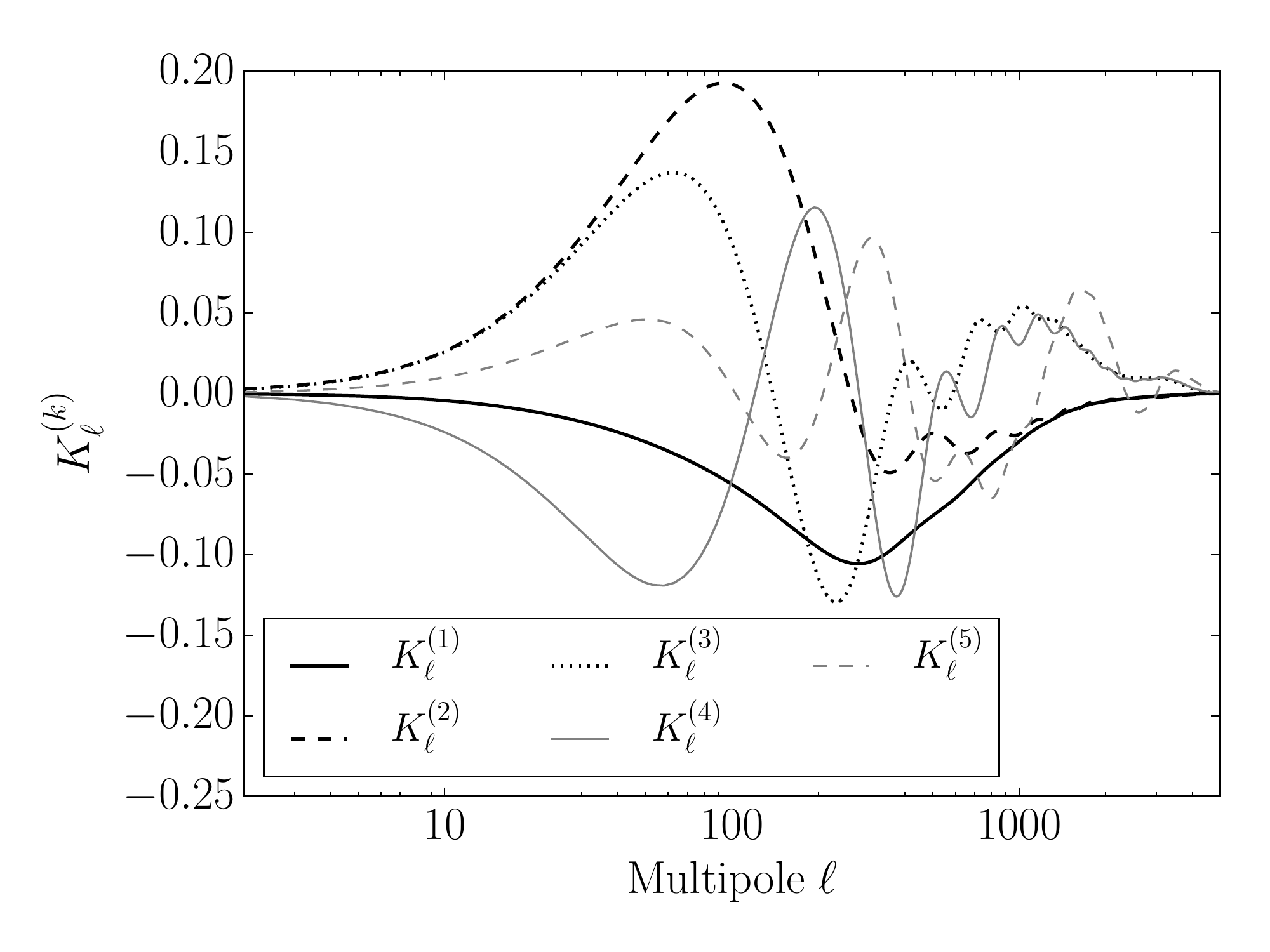}
\caption{Five principal components $K_\ell^{(i)}$ of the lensing potential best measured
by the lensed power spectra. 
}
\label{fig:pca}
\end{figure}

\subsection{Parameterizing lens and unlensed power spectra}

For a particular experiment and fiducial model, a Fisher forecast can determine which principal components
$\Theta^{(i)}$ of $C_\ell^{\PP}$ will be the best measured by the $XY$ power
spectra. These PCs can be ordered by increasing variance so that only the handful of
best measured components need be included in the actual analysis.
 In Ref.~\cite{Motloch:2016zsl}, the hierarchy of $\Theta^{(i)}$ was found for the experimental
configuration considered here. These principal components are defined as
\be
\label{Theta_definition}
	\Theta^{(i)} = \sum_{\ell} K^{(i)}_\ell ( \ln C^\PP_\ell   - \ln C_\ell^{\PP,{\rm fid}}),
\ee
{where}
$C^{\PP,{\rm fid}}_\ell$ is a fixed lensing potential, evaluated at the fiducial parameters 
given in Table~\ref{tab:fiducial}.   
The lensing PCs, $\Theta^{(i)}$, can be thought of as a more incisive generalization of the
standard approach where a scaling parameter $C_\ell^\PP \rightarrow A_L
C_\ell^\PP$ is added to test consistency of a model with lensing; PCs parametrize 
the lensing information more completely. 
For reasons detailed in
Appendix~\ref{sec:n_pca}, the 5 best measured PCs suffice for the data and models
considered here; their weights $K^{(i)}_\ell$ are shown in
Fig.~\ref{fig:pca}.\footnote{Note that in Fig.~6 of \cite{Motloch:2016zsl}, $K^{(i)}_\ell$
were scaled for display purposes.} 

 Moreover, the PCs decouple the
information on the lensing power spectrum from the parameters that control the unlensed
spectrum, whereas $A_L$ multiplies a $C_\ell^\PP$ that depends on such
parameters. PCs can
therefore be more directly compared with other measurements of $C_\ell^\PP$, most notably
from lensing reconstruction {using} the higher point information in the temperature and polarization fields themselves.

For example, in a cosmology where the unlensed
power spectra fluctuate low but the lensing potential is consistent with the
underlying model, there will be a tendency for high $A_L$: the unlensed CMB will drive the
amplitude of fluctuations $A_s$ down, which will at the same time lower the amount of
lensing predicted. To match the amount of lensing present in the data, $A_L$ or similar
lensing parameter is then increased.   Hence measuring
a high $A_L$ does not necessarily signal a deviation in $C_\ell^\PP$ itself.

In the PC approach, the unlensed CMB is described by separate parameters $\tilde\theta_A$.
For models that modify only the low redshift physics involved in the growth of structure and 
cosmic acceleration it is sufficient to take these to be the equivalent of the six $\Lambda$CDM parameters.  
These $\tilde \theta_A$ change the unlensed power spectra in exactly the manner of the $\Lambda$CDM
parameters, but unlike those, they have no effect on $C^{\PP}_\ell$.

To summarize, CMB power spectra $C^{XY}_\ell$ 
can be 
effectively parameterized in terms of eleven parameters $\tilde \theta_A$ and
$\Theta^{(i)}$ in any model that deviates from $\Lambda$CDM only after recombination,
with $i={1-5}$ sufficing for the data and models we study.
 In this setup, $\Theta^{(i)}$ represent 
direct, optimally weighted, measurements of the lensing potential.

\begin{figure}
\center
\includegraphics[width = 0.49 \textwidth]{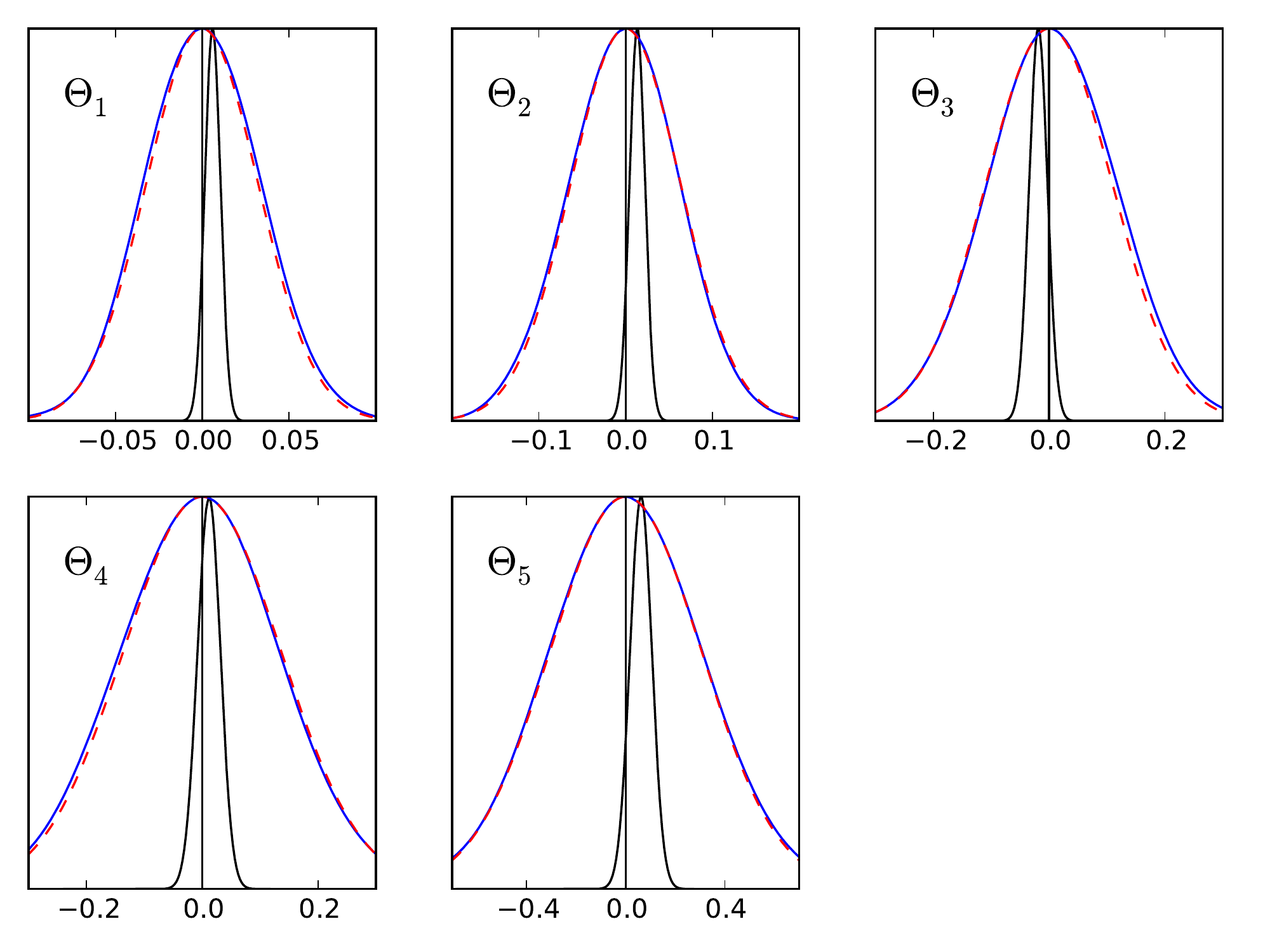}
\caption{Joint  posterior distribution  for PCs $\Theta^{(1-5)}$ (black) as a product of the individual 
posteriors from 50
independent all-sky
 simulations. 
The joint posterior is unbiased to a small fraction of the width of the distribution of
a single simulation (blue) and its Fisher prediction (red dashed).
}
\label{fig:theta_unbiased}
\end{figure}

\subsection{MCMC analysis of lens and unlensed parameters}
\label{sec:PCparams}

We run MCMC analyses on 50 independent lensed CMB sky simulations 
to check our likelihood model
and 
to determine properties of the parameters $\tilde \theta_A, \Theta^{(i)}$.

So far we have made numerous assumptions, for example that our models for the likelihood
and non-Gaussian covariance are correct, that the Fisher-based construction of the PCs
suffices, that neglecting higher $\Theta^{(i)}$ does not affect the constraints and that
agreement between {theoretical and
simulated power spectra} is sufficient (see Appendix \ref{sec:lenspix}).
It is thus a nontrivial check of our analysis to ascertain that the constraints on
$\Theta^{(i)}, \tilde \theta_A$ are unbiased with respect to the fiducial model.
To check this, we multiplied 50 MCMC posterior probabilities for $\Theta^{(i)}, \tilde
\theta_A$. The results for $\Theta^{(i)}$ are shown in
Fig.~\ref{fig:theta_unbiased} and show no significant bias relative to the standard deviation of a
single MCMC posterior; the same conclusion is valid also for $\tilde \theta_A$.

\begin{figure}
\center
\includegraphics[width = 0.49 \textwidth]{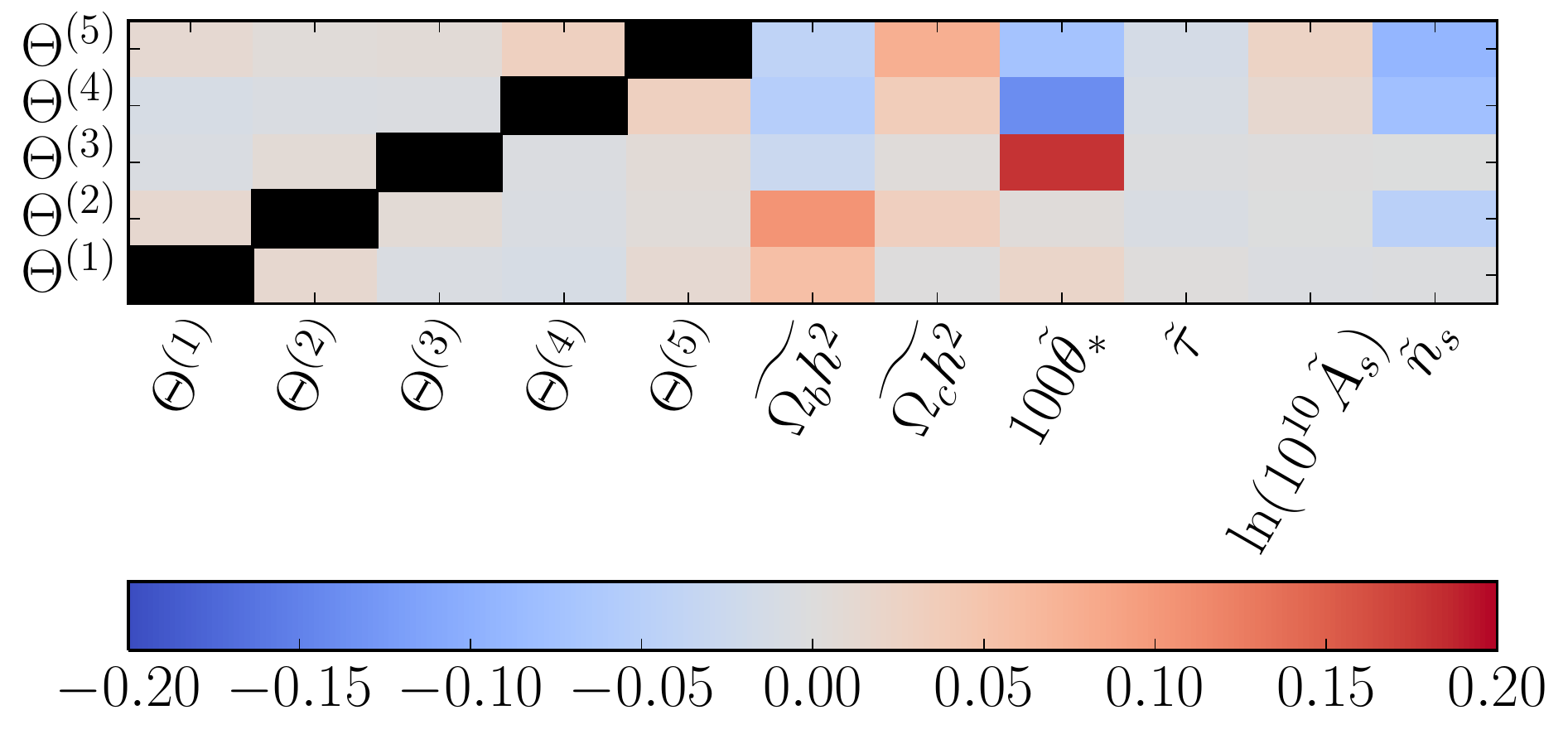}
\caption{Correlation matrix for $\Theta^{(i)}$ averaged over 50 MCMC analyses. 
{Black squares represent ones on the diagonal.}
The tilded parameters 
affect only the unlensed CMB, as explained in the text.
}
\label{fig:theta_uncorrelated}
\end{figure}

The Fisher analysis also predicts that $\Theta^{(i)}$ as determined by the data should be uncorrelated. This
assertion can be checked by averaging covariance matrix of the cosmological parameters
$\Theta^{(i)}, \tilde \theta_A$ over the MCMC analyses. Correlation coefficients 
of $\Theta^{(i)}$ obtained from this covariance matrix are shown in
Fig.~\ref{fig:theta_uncorrelated}. As expected, the lensing principal components are
only very weakly mutually correlated. Additionally, they
are also only mildly correlated with the unlensed parameters $\tilde \theta_A$. Most
significant of these are {$R=0.18$} correlation between $\Theta^{(3)}$ and
$\theta_*$ {and $R = -0.14$ correlation between $\Theta^{(4)}$ and $\theta_*$}.
This is somewhat counterintuitive, as $\theta_*$ shifts the angular scale of acoustic
features whereas lensing mainly smears the peaks by superimposing magnified and
demagnified regions. While largely true, the effect of lenses that are on scales smaller
than the acoustic scale $\ell \gtrsim 200$ is not purely a smearing effect, causing a
component that is slightly out of phase with the peaks in the unlensed power spectra,
leading to the observed correlation.

\begin{figure*}
\center
\includegraphics[width = 0.99 \textwidth]{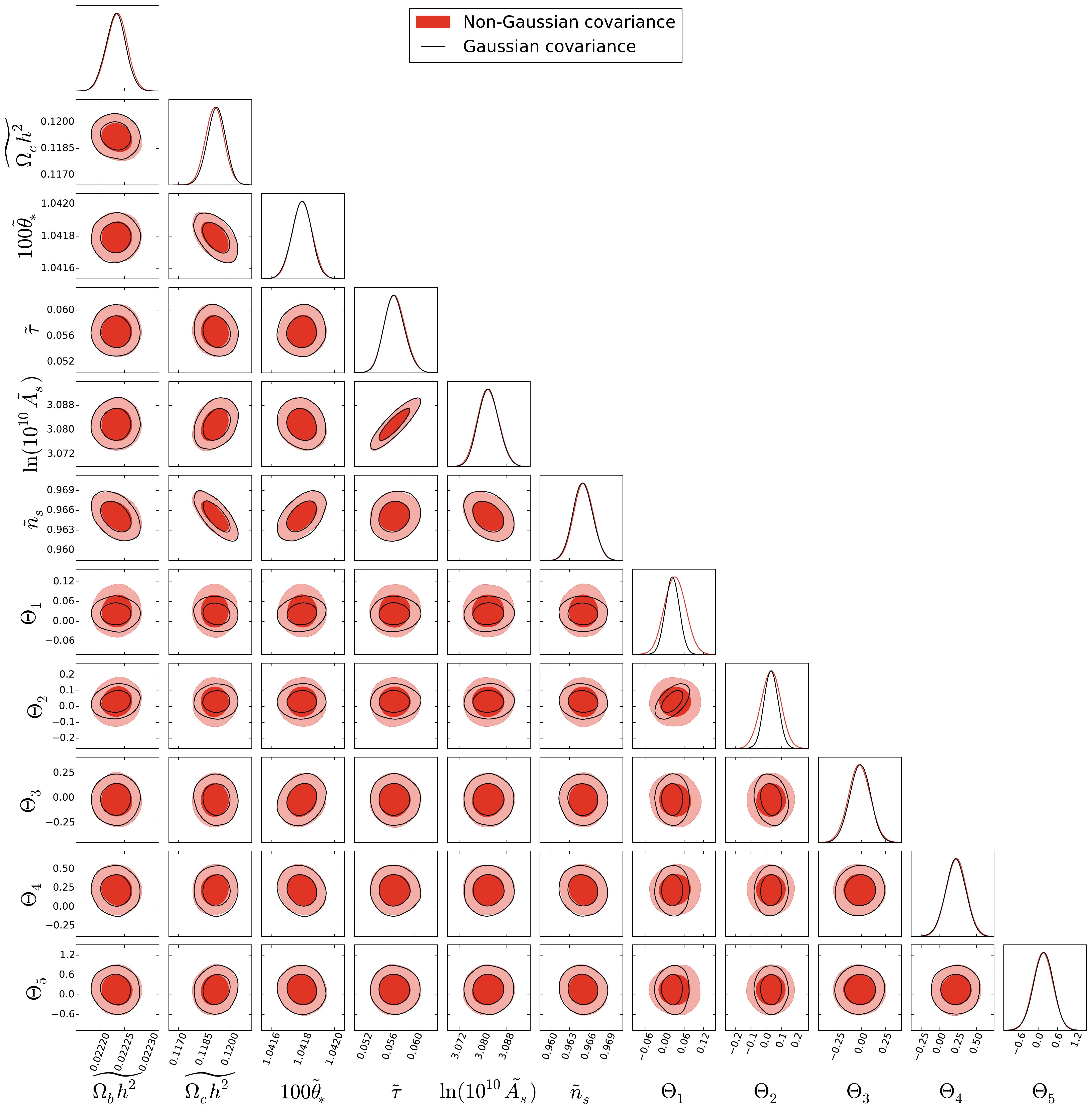}
\caption{MCMC constraints on the lens and unlensed parameters $\Theta^{(i)}, \tilde \theta_A$ in a typical simulation with a  Gaussian ({black} curves) and non-Gaussian covariance (red  shaded) analysis. 
}
\label{fig:lcdm+5}
\end{figure*}

The full results for all pairs of the 11 parameters in a single simulation are shown in Fig.~\ref{fig:lcdm+5}.  
Note that posterior distribution for the  parameters $\Theta^{(i)}, \tilde
\theta_A$ seems to be very well approximated by a multivariate normal distribution.
This implies that the Fisher approximation should be quite accurate in this space as we explicitly 
verify in Fig.~\ref{fig:theta_unbiased}.  We  exploit the multivariate normal nature of the
posterior in the $\Theta^{(i)}, \tilde
\theta_A$ variables in the following sections.

\subsection{Impact of lensing-induced covariance revisited}
\label{sec:explaining}

In Fig.~\ref{fig:lcdm+5} we also show the 
effect of the non-Gaussian covariance in the 11D lens and unlensed parameter space;
only $\Theta^{(1)}$ and $\Theta^{(2)}$ show significant effects of neglecting the
non-Gaussian likelihood. These two measurements are strongly affected, because significant
portion of the noise in these measurements arises from the sample variance of the lenses.
Neglecting lensing-induced terms in the data covariance effectively omits this noise
contribution, which leads to overly optimistic estimates on $\Theta^{(1)},
\Theta^{(2)}$. Measurements of other lensing principal components and $\tilde \theta_A$
are limited by other sources of noise -- instrumental noise and cosmic variance of the
unlensed CMB -- and the resulting constraints are thus not strongly affected by the non-Gaussian
part of the covariance.

If the eleven parameters $\Theta^{(i)}, \tilde \theta_A$
contain all information about a particular cosmological model, a Karhunen-Lo{\`e}ve
analysis applied to the Fisher information matrices {\cite{Smith:2006nk} } reveals
that non-Gaussian covariances can degrade the standard deviation of any linear combination
of these parameters by maximally 2.53.\footnote{{Using covariance matrices of
$\Theta^{(i)}, \tilde \theta_A$ from 10 MCMC simulations we checked this prediction is on
average correct to $\pm 0.04$.}} This generalizes the discussion of the most degraded
linear combination $M$ of cosmological parameters {from}
\S\ref{sec:cosmological_parameters}. This PC based quantification of the effect of
non-Gaussian covariances is not restricted to the models investigated here and can be
applied to more general extensions of $\Lambda$CDM. It is also not necessary to assume a
{\it linear} relationship between these effective parameters and the bare cosmological
parameters, or the validity of the Fisher approximation for the latter.

Moreover, we can use $\Theta^{(1,2)}$ to directly translate the effect of lensing-induced
covariance on cosmological parameter constraints.
Those combinations of cosmological parameters which are limited by our knowledge of
$\Theta^{(1,2)}$, in other words those constrained by the (mostly low $\ell$) lensing
information, will be strongly affected if we neglect the non-Gaussian covariance.

For example in $\Lambda$CDM, lensing information helps mainly $\omega_c$ and $A_s$
constraints. Increasing either of these parameters increases $C^{\phi\phi}_\ell$, lensing
information thus helps constrain the direction of simultaneously increasing
$\omega_c$ and $A_s$. As we can see in Fig.~\ref{fig:lcdm}, adding non-Gaussian
covariance weakens exactly this parameter combination the most. 
The reason becomes clear when we examine how change in the parameter combination $M$, the most sensitive to the non-Gaussian
covariances, projects onto the changes in the effective parameters $\Delta \Theta^{(i)},
\Delta \tilde \theta_A$. As expected, the main effect is a shift in  $\Theta^{(1)}, \Theta^{(2)}$, {which are the variables showing most of
the effect of the
non-Gaussian covariance; this shift is} captured in Fig.~\ref{fig:how_m_walks}. In
comparison, shifts in the other effective parameters $\Theta^{(i)}, \tilde \theta_A$ are
at least a factor of few smaller, as measured by the sizes of the marginalized posterior.
This means that already within $\Lambda$CDM it is possible to construct a parameter
combination which has a dominant effect of changing the lensing potential ($\Theta^{(1)},
\Theta^{(2)}$) and keeps the unlensed power spectra ($\tilde \theta_A$) relatively intact.
Because of that, the relative change in the standard deviation for $M$ brought about by the
non-Gaussian covariance 2.03 is already close to the maximal possible value of 2.53.

\begin{figure}
\center
\includegraphics[width = 0.49 \textwidth]{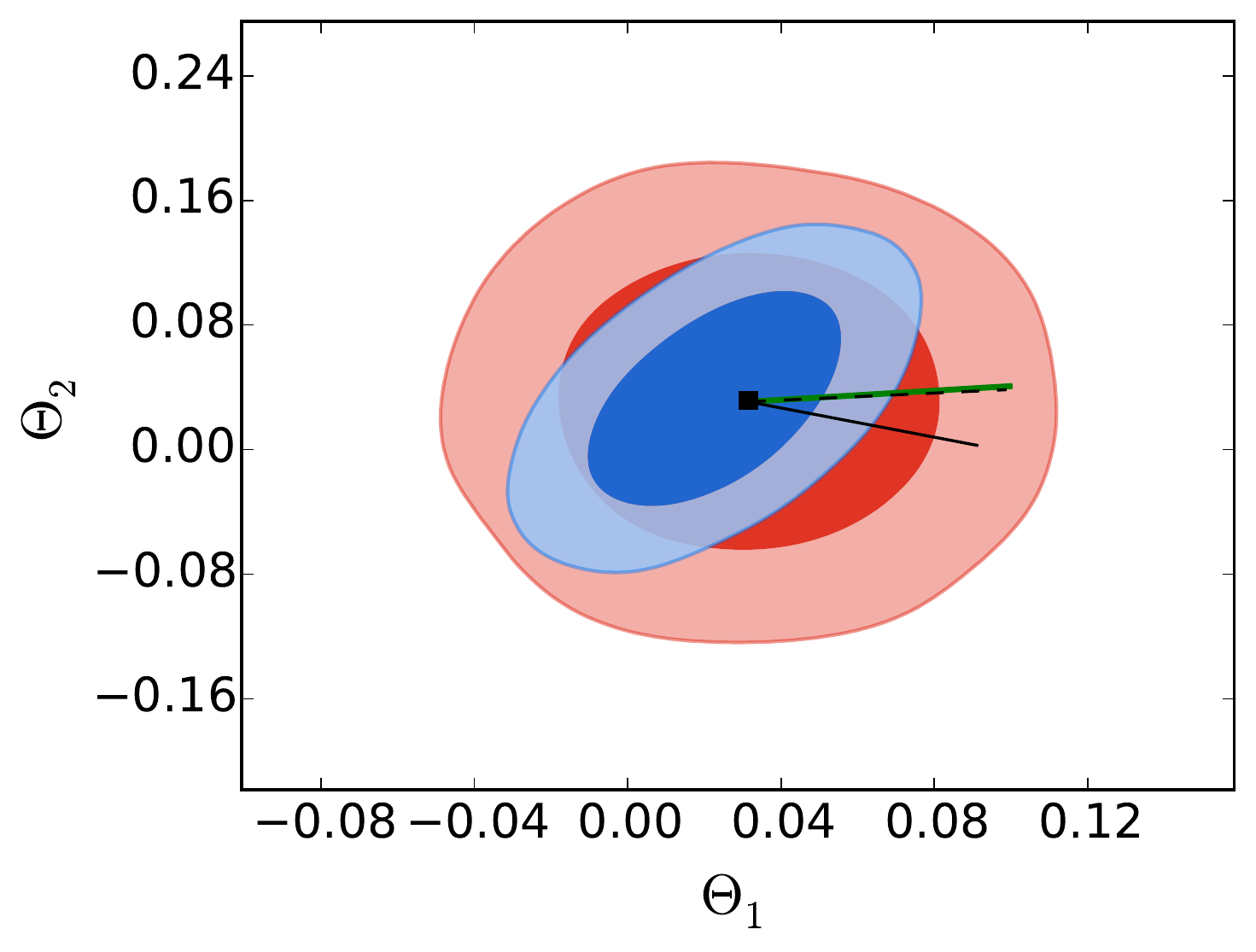}
\caption{Lines show how $\Theta^{(1)}$ and $\Theta^{(2)}$ 
change when we increase $M$ (black dashed), $M^\nu$ (green) or $M^w$ (black solid);
length of the lines is arbitrary. For comparison, in the background we show typical
constraints on these two parameters in an MCMC analysis based on non-Gaussian (red) and
Gaussian (blue) covariance.}
\label{fig:how_m_walks}
\end{figure}

By extending the cosmological model to $\Lambda$CDM+$\sum m_\nu$, we increase the
parameter freedom, which enables us to find a parameter combination which is slightly more
limited by the lens sample covariance and shows degradation of 2.05. 
From the perspective of the effective parameters $\Theta^{(i)}, \tilde
\theta_A$ this occurs because it is possible to achieve the same change of the lensing
potential, as seen in the nearly identical directions of $M$ and $M^\nu$ in Fig.~\ref{fig:how_m_walks}, with a smaller change
in the unlensed power spectra.  This increases relative importance of the low $\ell$
lensing information $\Theta^{(1)},
\Theta^{(2)}$ in constraining $M^\nu$, which directly leads to a larger impact of
non-Gaussian covariance.  

Moreover, lensing information is now important for three parameters
$\{ \omega_c, A_s, \sum m_\nu\}$ unlike two in $\Lambda$CDM.
 Because the degeneracy structure involves  three
parameters, the non-Gaussian effect is hidden from the covariance of any two, once the
third is marginalized. In a three-dimensional likelihood for these three parameters, the
effect of non-Gaussian covariance is clearly visible.

For $\Lambda$CDM+$w$ the analysis is similar. Again, by adding a new parameter on top of
$\Lambda$CDM we can find $M^w$ which shows degradation larger than what is seen in
$\Lambda$CDM, in this case 2.28. For $\Lambda$CMD+$w$ this happens, because the
projection of $M^w$ onto  $\{  \Theta^{(1)},  \Theta^{(2)}\}$ is more
aligned with the
direction maximally impacted by the non-Gaussian covariance, see Fig.~\ref{fig:how_m_walks}.

\subsection{Consistency tests}

The simplest lensing consistency test of a model is to compare lensing potential measured through $\Theta^{(i)}$ against
what is expected based on the cosmological model determined by constraints on the unlensed
power spectra $\tilde \theta_A$.   This test mainly checks the internal 
consistency of the model assumptions. 

If there are also additional measurements of the lensing potential,  then a sharper
consistency check is possible \cite{Motloch:2016zsl}.  A certain linear
combination of the power spectrum PCs,
\be
\label{psi1_combination}
	\Psi^{(1)} = \sum_{i = 1}^{5} \mathcal{T}_i \Theta^{(i)}
\ee
with $\mathcal{T}_i =
\{ 
32.3,
-15.9,
0.330,
1.72,
-0.608
\}$,
is predicted by the Fisher analysis to be limited mainly by lens sample variance, which 
drops out when comparing with other measurements on the same
patch of sky leaving a nearly noise-free consistency test.   This consistency test therefore checks for systematics in the data analysis,
foregrounds, and assumptions about the unlensed power spectra. 

In Fig.~\ref{fig:psi1_scatter}, we illustrate this idea explicitly by comparing
{posterior} {mean} values of $\Psi^{(1)}$ determined from 50 simulated power spectra using MCMC
against values of $\Psi^{(1)}$ determined directly from a known realization of the lensing
potential, which can be thought of as a limiting case of lensing potential measurement
with no instrumental noise. The latter approach estimates $\Theta^{(i)}$ from simulated
$C_\ell^{\phi\phi}$ using an unbiased estimator
\be
\label{Theta_definition}
	\Theta^{(i)}_\mathrm{est} = \sum_{\ell} K^{(i)}_\ell ( \ln C^\PP_\ell   - \langle\ln C_\ell^{\PP}\rangle),
\ee
where $\langle \cdot \rangle$ is expectation value over realizations, and combines the
results according to \eqref{psi1_combination}. The observed correlation is indeed very
tight; residual scatter in Fig.~\ref{fig:psi1_scatter} is caused by instrumental noise and
variance of the unlensed CMB which affect the power spectrum measurement. 

\begin{figure}
\center
\includegraphics[width = 0.49 \textwidth]{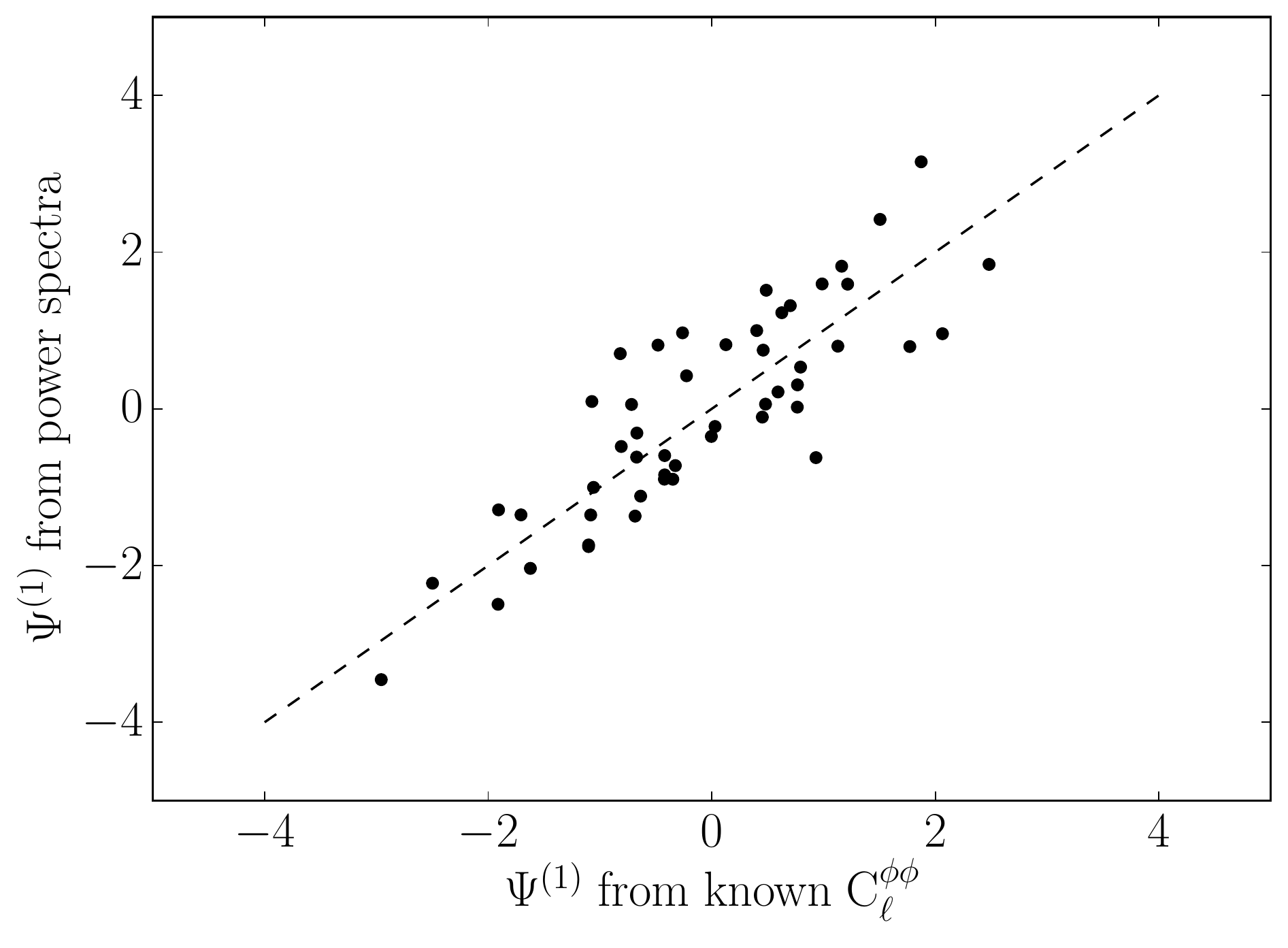}
\caption{Consistency mode $\Psi^{(1)}$ as determined from {posterior} {mean} values in
50 simulated lensed power spectra through MCMC analysis against values determined from
known realizations of the lensing potential (see text for details). The dashed line
represents points where {the} two {determinations} are equal.
}
\label{fig:psi1_scatter}
\end{figure}

Notice also that this direction is mainly composed of $\Theta^{(1)}$ and $\Theta^{(2)}$ 
in nearly the same combination that is maximally affected by non-Gaussian lens sample variance.
{While being limited by the lens sample variance is detrimental to cosmological
parameter constraints, for consistency tests this is advantageous because lens sample
variance drops out when comparing measurements on the same patch of sky.}

There is also another combination of PCs,
$\Psi^{(2)}$,
\be
\label{psi2_combination}
	\Psi^{(2)} = \sum_{i = 1}^5 \mathcal{U}_i \Theta^{(i)}
\ee
with $\mathcal{U}_i =
\{ 
31.3,
17.6,
10.4,
0.945,
-1.10
\}$,
which can serve as a similar consistency check; this consistency check is
slightly weaker than the test using $\Psi^{(1)}$ due to the larger impact of noise and
sample variance of the unlensed CMB. {This weakening can be seen in
Fig.~\ref{fig:psi2_scatter}, where we compare values of
$\Psi^{(2)}$ determined from 50 simulated power spectra against values from known realization of the
lensing potential.}

\begin{figure}
\center
\includegraphics[width = 0.49 \textwidth]{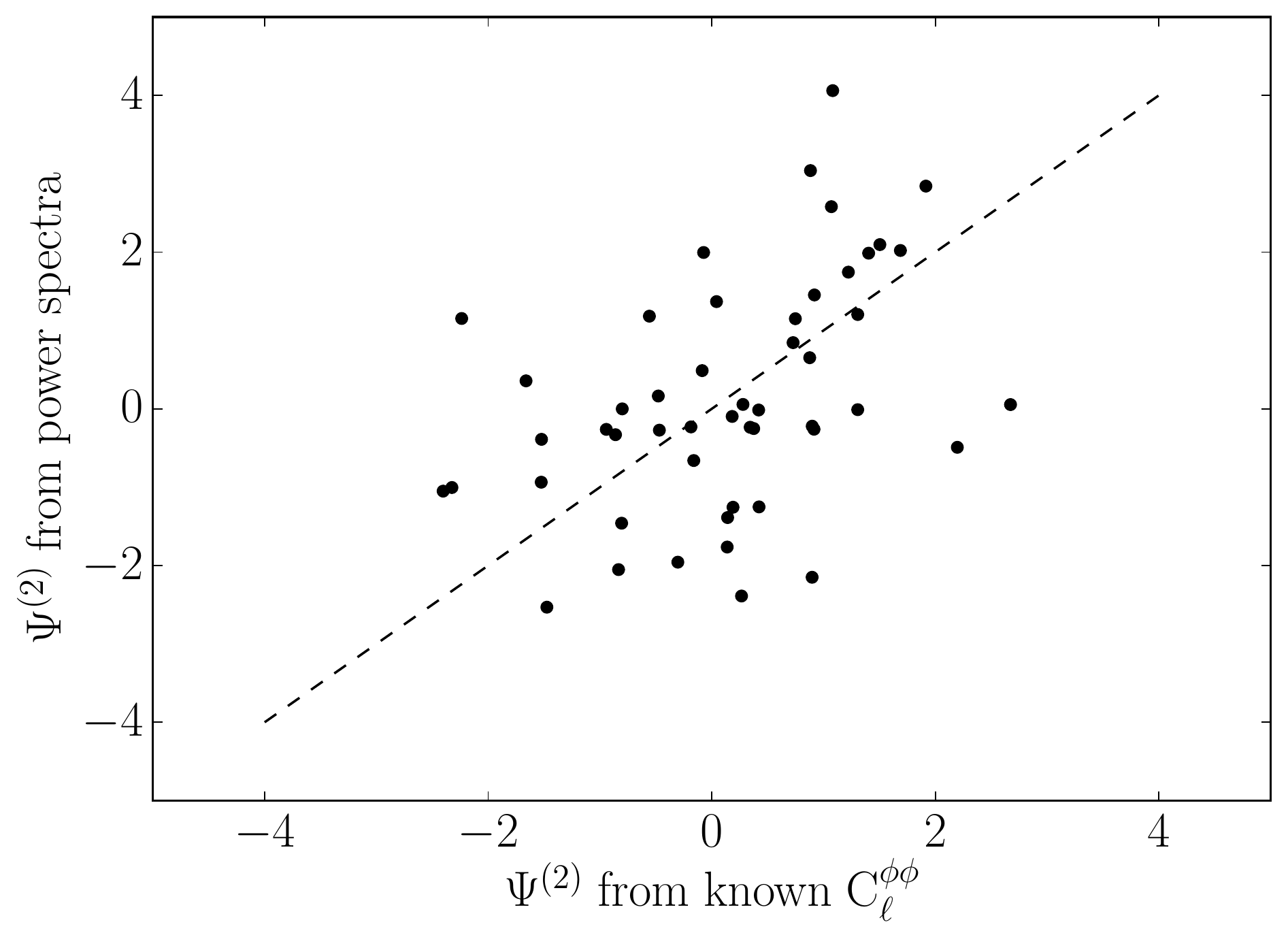}
\caption{
Same as Fig.~\ref{fig:psi1_scatter} but for the consistency mode $\Psi^{(2)}$.
}
\label{fig:psi2_scatter}
\end{figure}

Notice also that there is no indication of bias in the power spectrum estimate of the two consistency modes in 
Figs.~\ref{fig:psi1_scatter}, \ref{fig:psi2_scatter}.   
To  better quantify this, 
in Figure~\ref{fig:psi_unbiased} we show product of 50 posteriors for
measurements of $\Psi^{(1)}, \Psi^{(2)}$ as determined from our MCMC simulations and the determination
is indeed unbiased; from the power spectra side there does not seem to be any problem for
the consistency check.

\begin{figure}
\center
\includegraphics[width = 0.49 \textwidth]{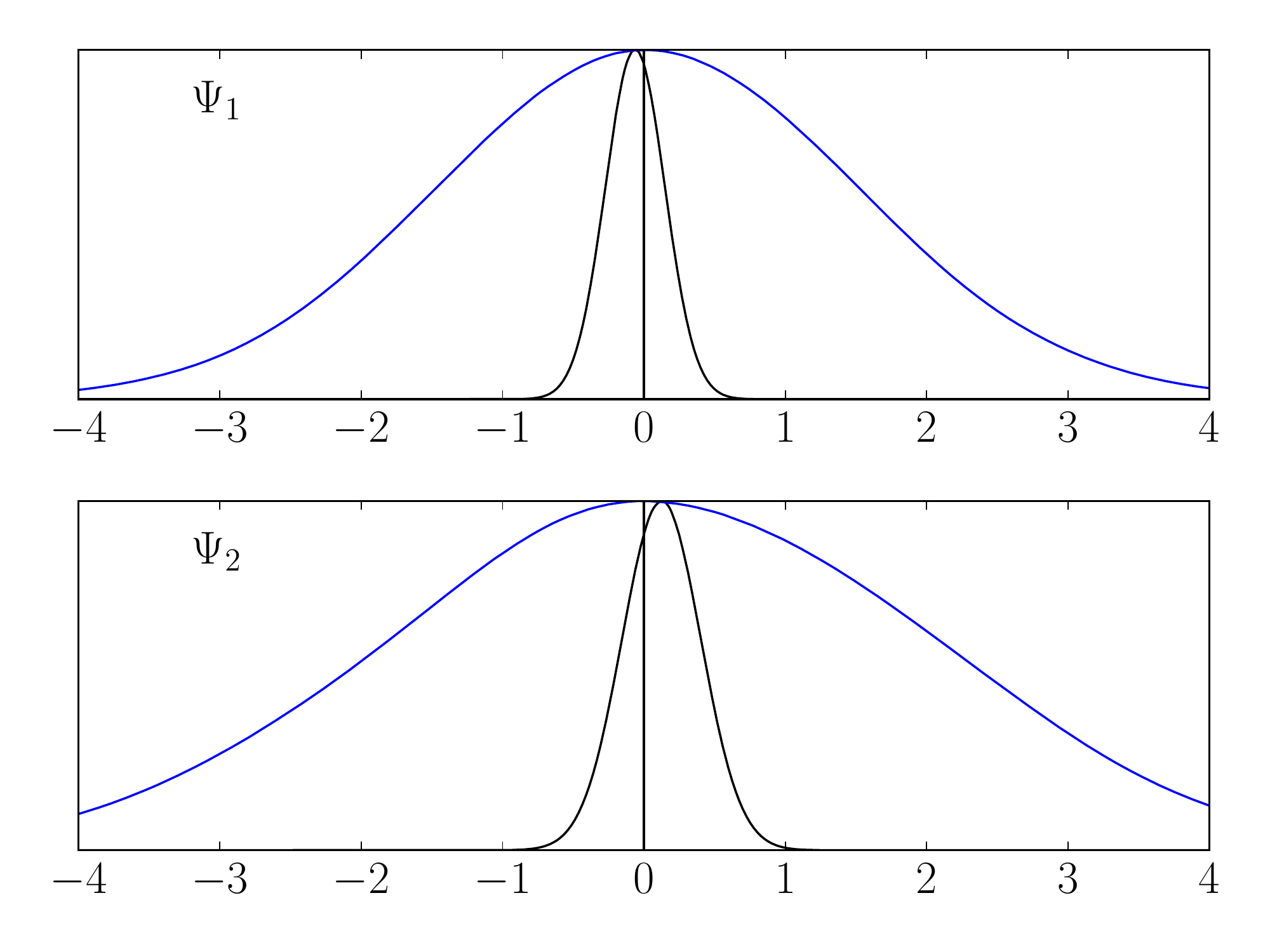}
\caption{Joint posterior of the consistency {parameters $\Psi^{(1)}, \Psi^{(2)}$} (black) as the product of 50
individual posterior distributions from independent all-sky simulations.
 Compared against the {width of} a single posterior ({blue}) there is no indication of bias 
with respect to the
fiducial value $\Psi^{(i)} = 0$ at a fraction of the standard deviation.  }
\label{fig:psi_unbiased}
\end{figure}

We also find that $\Psi^{(1)}$ is almost uncorrelated with the $\Lambda$CDM
parameters describing the unlensed power spectra; the largest correlation coefficient we
find is $R = -0.03$ and occurs between $\Psi^{(1)}$ and $\widetilde {\Omega_b h^2}$. The other
consistency mode is slightly more correlated with the unlensed parameters; the
most correlated with $\Psi^{(2)}$ are $\tilde \theta_*$ with correlation coefficient $R =
0.12$ and $\widetilde {\Omega_b h^2}$ with $R = 0.08$.

Failure of the consistency check could indicate an unlensed spectrum that is not described
by the flat $\Lambda$CDM parameters.   In fact in \cite{Motloch:2016zsl} it was shown that
the second consistency mode $\Psi^{(2)}$ is correlated with the spatial curvature given
their similar effects on the acoustic peaks. We run a single MCMC analysis in which we
added $\tilde \Omega_K$ to the unlensed parameters 
and confirm this finding; there is a strong correlation between $\tilde \Omega_K$ and
$\Psi^{(2)}$ with correlation coefficient $R = -0.62$. In a non-flat Universe analyzed as flat
this would lead to failure in the consistency check as $\Psi^{(2)}$ would move from its
true value to absorb the unaccounted for curvature. The main consistency mode, $\Psi^{(1)}$, is correlated with
$\tilde \Omega_K$ at the $R = 0.29$ level and is a weaker check on curvature.

\subsection{Effective Likelihood for Model Building}

Given the nearly multivariate normal posterior probability of  
the effective parameters $d_a = \{\tilde \theta_A , \Theta^{(i)}\}$,
we can also use a single MCMC analysis to  compress the whole CMB power
spectra data into  11 numbers for the mean {values $\bar d_a$} and the $11\times 11$ covariance matrix of $d_a$.
These can be used to form an
effective likelihood function ${\cal L}_{\rm eff}(\bar d_a |\theta_A)$
{defined as}
\begin{equation}
-2\ln{\cal L}_{\rm eff} =\sum_{ab} [ \bar d- d(\theta_A)]_a ({\rm Cov}_{d}^{-1})_{ab} [\bar d- d(\theta_A)]_b .
\end{equation}
Here {$d_a(\theta_A$)} models the expectation values for the data $\bar d_a$
{as a function of the cosmological parameters $\theta_A$ of a given cosmological
model.}
This effective likelihood can be now used to probe a broad
class of {cosmological} models without any explicit use of the raw CMB power spectra data {by
specifying $d_a(\theta_A)$ for each such model}.
{Class of models for which this approach is effective} contains not just
{$\Lambda$CDM+$w$ and $\Lambda$CDM+$\sum m_\nu$} considered here but
also models which are indistinguishable from $\Lambda$CDM at recombination and
for which CMB lensing is the dominant source of information on the {physics beyond
$\Lambda$CDM}.   For
example, many models of dark energy and modified gravity fall into this class, if we are
willing to ignore the extra information coming from the integrated Sachs-Wolfe effect and
other secondaries.  In principle, the technique can be extended to incorporate such effects by 
extending the set of unlensed parameters $\tilde \theta_A$.

In the context of model building, one can envision a scenario where $\Lambda$CDM parameters
produce a poor effective likelihood for the data and motivate {extensions beyond
$\Lambda$CDM}.    The
effective likelihood can then be used as a quick spot check as to whether the given extension improves
the fit.

Let us illustrate this technique on {$\Lambda$CDM+$w$ and $\Lambda$CDM+$\sum m_\nu$}. 
{First, it} is necessary to find the functional dependence of $d_a$ on the
cosmological parameters $\theta_A$.   The values of the unlensed parameters $\tilde \theta_A$ 
{for
$ A \in \{{\theta_*},
{\Omega_c h^2},
{\Omega_b h^2},
{n_s},
{{\ln A_s}},
{\tau}
\}$}
are 
the same as the true cosmological parameters $\theta_A$,
while {the} values of the lensing principal
components  $\Theta^{(i)}$
can be determined directly from the definition \eqref{Theta_definition} given
$C_\ell^{\phi\phi}$ alone.
The full parameter space of the given extension can {then} be explored with an MCMC 
in the general case where $d_a(\theta_A)$ is nonlinear across the allowed region of the
parameter space as in the $\Lambda$CDM+$w$ extension.
  In a case such as
$\Lambda$CDM+$\sum m_\nu$, where the mapping can be linearized, it is possible to get a
good estimate of parameter constraints even without performing any additional MCMC.

In Fig.~\ref{fig:mi_lcdm+w} we show comparison of $\Lambda$CMD+$w$ parameter constraints
obtained in the analysis presented in the previous paragraph against results of the
standard MCMC analysis.
Because the mapping onto the effective parameter space is non-linear, it is necessary to
perform an additional MCMC run.  Note that this mapping alone accounts for most of the
non-normal posterior probability in the $(\omega_c,w)$ plane despite being based on a normal 
distribution for the effective parameters.   It slightly underestimates the lower limit on $w$,
presumably due to the neglect of the integrated Sachs-Wolfe effect in the unlensed parameters.
 For $\Lambda$CDM+$\sum
m_\nu$, the agreement between the simplified and standard analyses is even better.

\begin{figure}
\center
\includegraphics[width = 0.50 \textwidth]{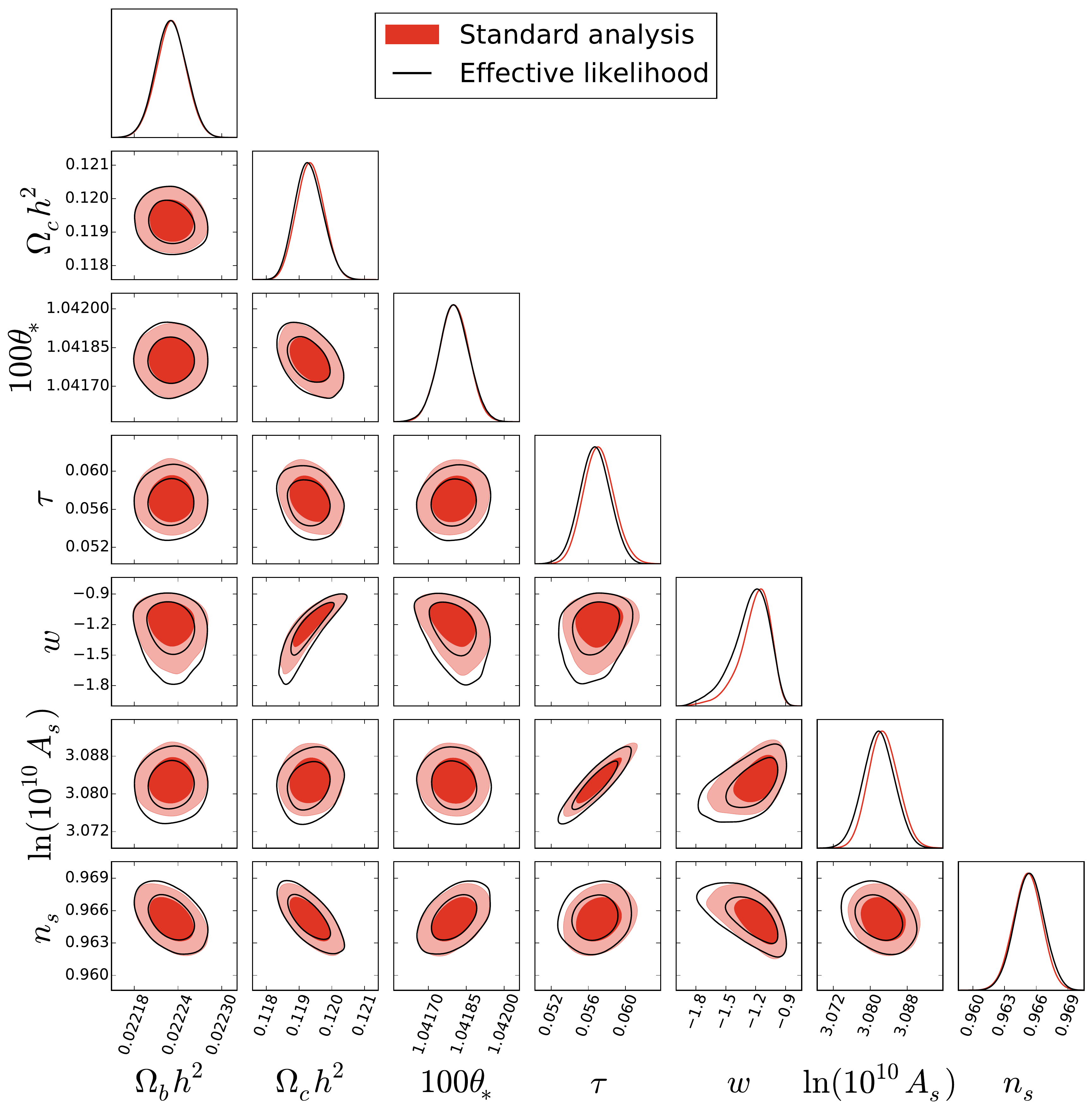}
\caption{Comparison of constraints on $\Lambda$CDM+$w$ parameters from the standard
analysis (red shaded) with results of an approximate analysis based on the effective
likelihood of $\{\Theta^{(i)},\tilde \theta_A\}$ instead of the raw CMB data ({black}).
}
\label{fig:mi_lcdm+w}
\end{figure}

\section{Discussion}
\label{sec:discuss}

Future measurements of lensed CMB power spectra will be increasingly affected by the lens
sample variance and its effect on parameter constraints will have to be included into the
analysis pipelines once polarization measurements approach the sample variance limit.
To this end we have developed and tested such an analysis pipeline starting from simulated lensed maps with CMB-S4 level instrument noise through to constraints on  cosmological parameters and the lens 
power spectrum.

The first piece in the pipeline is  a model for the likelihood function of full-sky lensed CMB power spectrum data, which
includes the non-Gaussian effects of lensing sample variance that correlate the measurements. This model considers large and small multipole
data as independent.    At small multipoles, the
likelihood assumes data are Wishart distributed and drops the small non-Gaussian effects of lensing, while at large multipoles it
assumes they are multivariate-normal distributed, with a covariance matrix which
includes the lensing correlation across multipole moments. 

With this likelihood we investigated parameter constraints from simulated lensed CMB
maps of a fiducial $\Lambda$CDM model.  We obtain MCMC parameter constraints on
the  $\Lambda$CDM parameters as well as two extensions,
$\Lambda$CDM+$\sum m_\nu$ and $\Lambda$CDM+$w$. The dominant effect of the
lensing-induced covariance in all of the models is more than four-fold increase in
variance of particular combinations of cosmological parameters $M,M^\nu,M^w$. As a
consequence, if the analysis of the real data was performed using Gaussian covariance in
the likelihood, instead of the proper non-Gaussian covariance, there is a high chance of
committing Type 1 error - mistakenly ruling out true cosmological model. This would potentially
affect concordance studies comparing constraints from various datasets. Shifts in the best
fit basis parameters and change in constraints of the other parameter combinations are typically
not as significant due to marginalization. 
The exception is $\Lambda$CDM+$w$ where a significant degradation in the
lower limit for $w$ is manifest in the MCMC results.   This degradation is hidden from the local Fisher forecasts as well as previous studies 
due to the strongly non-normal posterior distribution of $w$.

Then we explored the use of direct constraints on the lensing potential through a principal
component analysis.  Here eleven parameters effectively describe most of the
cosmological information contained in the lensed CMB power spectra. Five of these parameters,
$\Theta^{(i)}$, are  the best measured principal components of the lensing potential while the
remaining six, $\tilde \theta_A$, parameterize the unlensed power spectra. Measurement of $\Theta^{(i)}$ from data are well suited for various consistency tests involving measurements of the lens power
spectrum. This should be contrasted with the standard approach where cosmological parameters are augmented by a
scaling parameter $A_L$ to the lens power spectrum but the latter
 itself depends on cosmological parameters which are subsequently marginalized.  
 Here the measured $\Theta^{(i)}$ can be compared directly
 against lensing potentials corresponding to the measured unlensed parameters
$\tilde \theta_A$, to check the internal consistency of a particular cosmological model

On 50 MCMC analyses we tested our PC analysis pipeline, not finding any
significant bias in either $\Theta^{(i)}$ or $\tilde \theta_A$. The lensing
principal components $\Theta^{(i)}$ are found to be only weakly correlated, both mutually
and with the unlensed parameters $\tilde \theta_A$. The majority of the effects of the
non-Gaussian covariance consists of degrading constraints on the two leading lensing
principal components, $\Theta^{(1)}$ and $\Theta^{(2)}$ (see Fig.~\ref{fig:lcdm+5}).
This allows explanation of the parameter constraint degradations seen in the cosmological
models - in each there is a parameter combination which is predominantly limited by the
low $\ell$ lensing information which has large lens sample variance relative to other
sources of noise. Neglecting
non-Gaussian terms in the covariance effectively neglects this source of noise, which
misleads the parameter constraints.

The effect of lens sample variance on the PCs  enables a
sharp consistency test against other measurements of the lens power spectrum.
The degradation in parameter errors reflects a linear combination of PCs whose measurements are nearly lens sample variance limited.   Independent measurements on
the same patch of sky, e.g. through direct reconstruction from the CMB four point functions,
should agree since the sample variance is common to both. In this paper we
checked that the combinations of $\Theta^{(i)}$ which are expected to form the most
stringent consistency tests, $\Psi^{(1)}$ and $\Psi^{(2)}$, satisfy theoretical
expectations. Their values determined from lensed power spectra using MCMC analyses are
correlated with the ``true'' values determined from the known realization of $C_\ell^\PP$.
They are also unbiased and {nearly} uncorrelated with the unlensed parameters
$\tilde \theta_A$.

Failure of any of these consistency tests may indicate new physics beyond flat $\Lambda$CDM,
as we show on an example of spatial curvature.  
Constraints on $\Theta^{(i)}, \tilde \theta_A$ can also be used for model building purposes
given their simple multivariate normal form.   As illustrated using the dark energy equation of state,  one can rapidly explore lensing constraints on extensions to $\Lambda$CDM
using an effective likelihood for these parameters without recourse to the original CMB power spectrum data or experimental specifics.

\acknowledgements{
We thank Chen He Heinrich, Alessandro Manzotti and Marco Raveri for useful discussions.
This work was
supported by NASA ATP NNX15AK22G and the Kavli
Institute for Cosmological Physics at the University of Chicago through grant NSF
PHY-1125897 and an endowment from the Kavli Foundation and its founder Fred Kavli.  WH was
additionally supported by   U.S.~Dept.\ of Energy contract DE-FG02-13ER41958  and the Simons Foundation.  
PM acknowledges the hospitality of Nordita where part of this work was completed 
during  its program Advances in Theoretical Cosmology in Light
of Data.
We acknowledge use of the CAMB, Lenspix and CosmoMC software packages. 
This work was completed in part with
resources provided by the University of Chicago Research Computing Center.
}

\appendix

\begin{figure}
\center
\includegraphics[width = 0.49 \textwidth]{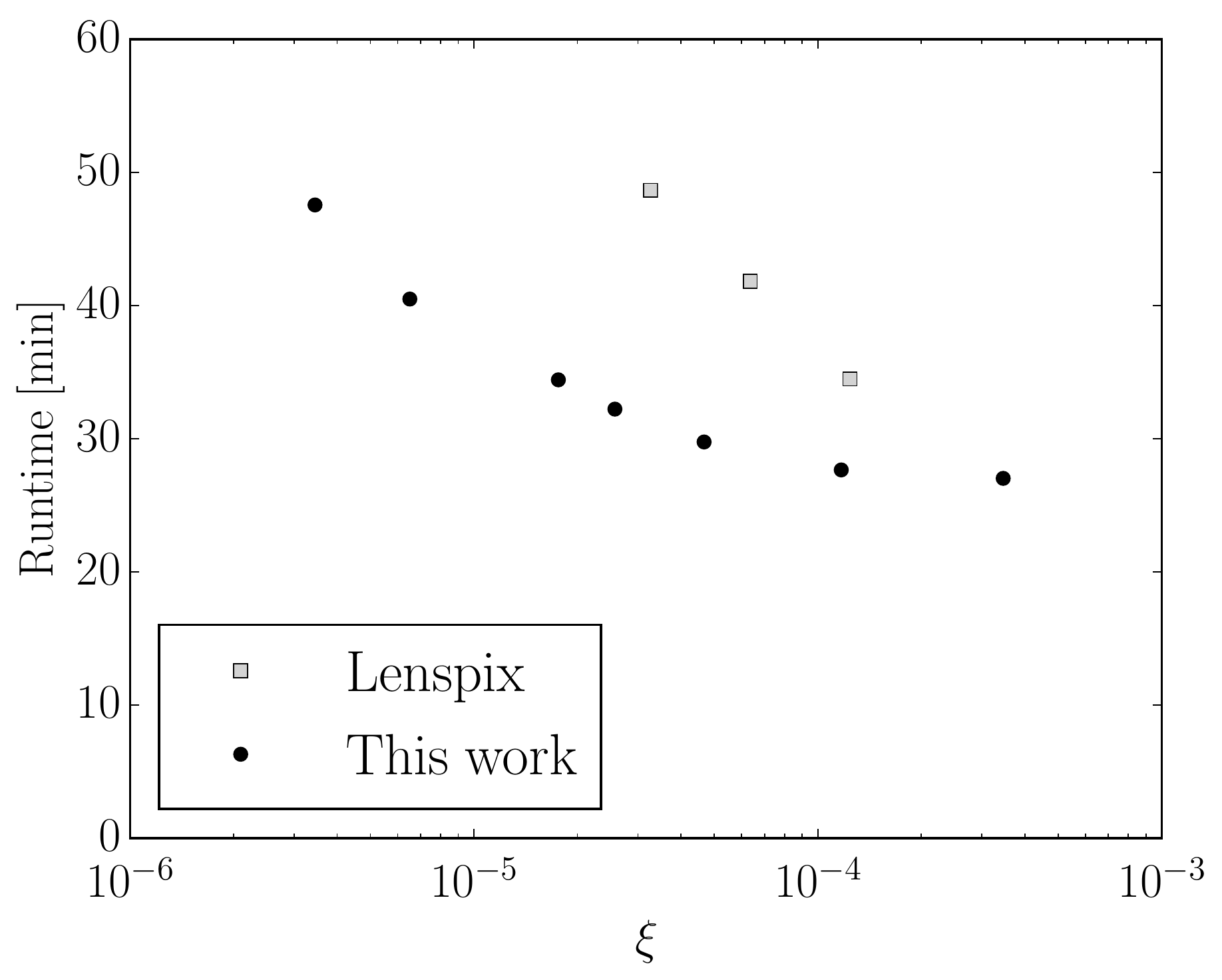}
\cprotect\caption{Bias $\xi$ caused by the interpolation part of the lensing algorithm and
corresponding runtime for various values of the precision parameter
\verb|interp_factor|. Comparison of the Lenspix interpolation routine (gray squares,
from right \verb|interp_factor| values 2, 2.5 and 3) and our modifications 
(black dots, values 2,
2.5, 3, 3.5, 4, 5 and 6).
} 
\label{fig:lenspix_comparison} 
\end{figure}

\section{Simulated CMB sky and Lenspix modifications}
\label{sec:lenspix}

To simulate lensed CMB data we modify publicly available code Lenspix\footnote{
https://github.com/cmbant/lenspix} \cite{Lewis:2005tp}. In this code the unlensed CMB is
first evaluated on a high resolution equicylindrical grid. The lensed CMB is then evaluated on
a lower resolution Healpix grid \cite{Gorski:1999rt} through a remapping by a deflection
field, determined by a gradient of the lensing potential $\phi$. Values of the unlensed CMB at
points which are remapped onto the Healpix grid points are obtained using a bi-cubic
interpolation from the high resolution grid.  Our simulations are run with precision
parameters \verb|nside| = 4096 and \verb|lmax| = 8000.

The precision with which the code calculates the lensed power spectra depends on the point density
of the high resolution grid, which is parameterized by an oversampling factor
\verb|interp_factor|. Simulations with \verb|interp_factor| $\sim$ 2, which were the
largest we could originally run on a single node of our computer cluster due to limited memory, lead to lensed CMB
power spectra biased at high $\ell$. Such bias leads to $\sim 0.2$ standard deviations
shift in the likelihood function in the $\ln A_s$ direction (with other
cosmological parameters fixed); we did not investigate other
parameters in depth but in general parameters constrained by high $\ell$ data are
sensitive to this bias.

This power spectra bias can be quantified by a parameter 
\be
	\xi = \frac{C^{TT}_{3000}\Big|_{\phi = 0} - \tilde C^{TT}_{3000}}{\tilde C^{TT}_{3000}} ,
\ee
relative difference at $\ell = 3000$ of the temperature power spectrum $C^{TT}$ 
``lensed'' in Lenspix by a zero-deflection field and the unlensed temperature power spectrum
$\tilde{C}^{TT}$. If interpolation was exact $\xi$ would vanish.
However, the
unlensed and lensed CMB are evaluated at different grids and interpolation leads to
numerical bias even when there is no lensing present. This bias appears to be -- up to
cosmic variance -- independent of the cosmology and comparable in temperature and
polarization. It typically grows with investigated multipole, for comparison we therefore choose
the largest data multipole considered in the paper, $\ell = 3000$. 

To overcome the large $\xi$ bias and avoid the related shifts in the likelihood function, we
modify the code such that it works only with smaller portions of the high resolution
map of unlensed CMB at any given time and never stores the whole map in memory. This
allows us to run with higher values of \verb|interp_factor| and achieve smaller values of
$\xi$.

We further replace the original high precision calculation of partial
derivatives of the unlensed CMB variables, which is part of the Lenspix interpolation algorithm, by
a less precise (for a given high resolution grid) but significantly faster routine. This
enables us to obtain higher interpolation precision without sacrificing runtime by
increasing the density of the high resolution grid of the unlensed CMB. 

Finally, the precision of variables describing the angular positions of the points in the high
resolution grid is increased to avoid certain artifacts in lensed CMB maps. 

We compare values of $\xi$ and runtime for several
values of \verb|interp_factor| with the original and simplified calculation of the partial
derivatives in Fig.~\ref{fig:lenspix_comparison}. It is clear that although the original
routine is superior for a fixed high resolution grid, for a fixed runtime it is
advantageous to use a simpler partial derivative calculation and increase the density of
the grid. Simulations used in this work were calculated with \verb|interp_factor|
{= 4.}

\begin{figure}
\center
\includegraphics[width = 0.49 \textwidth]{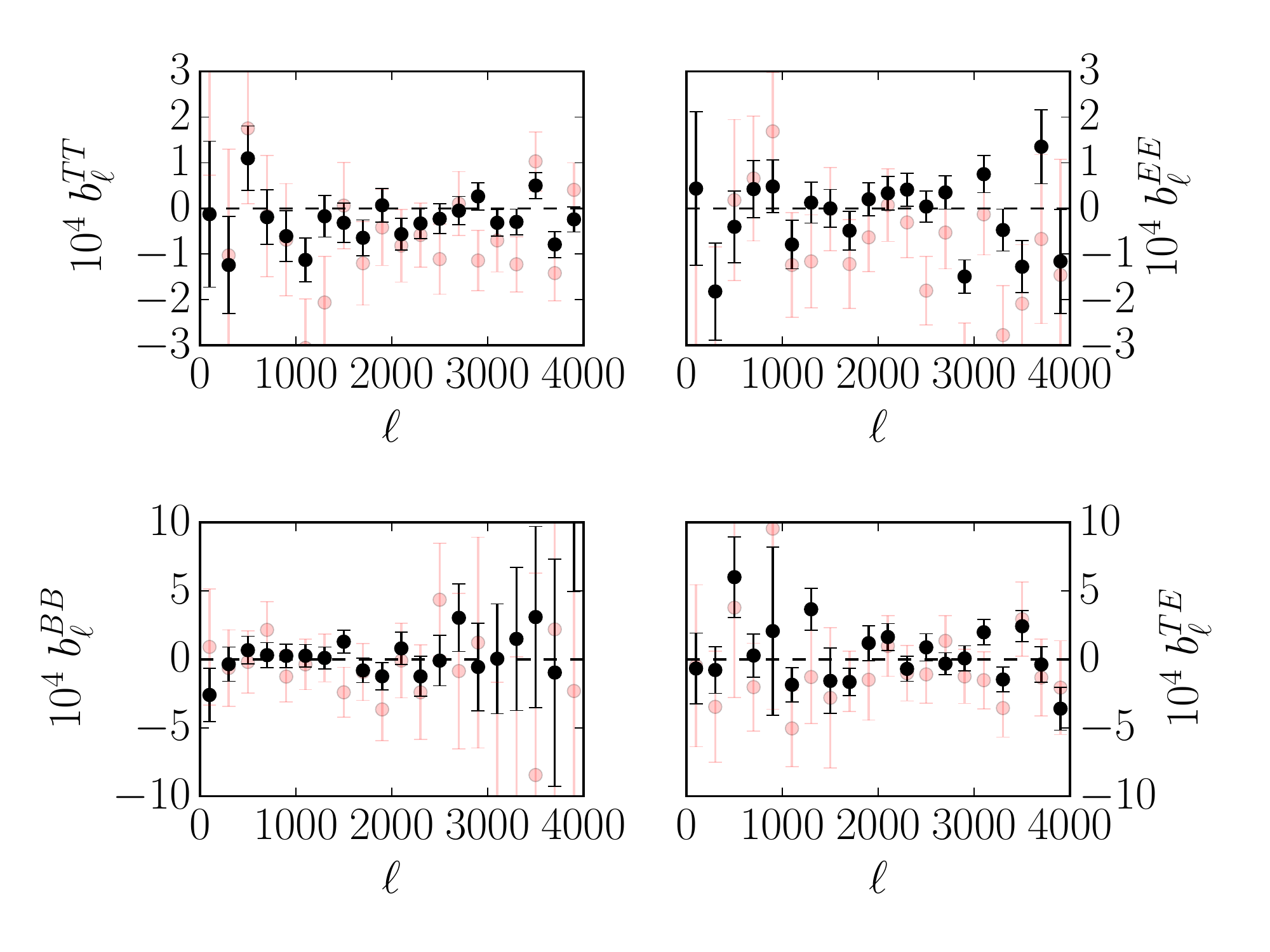}
\cprotect\caption{{Lensed power spectra bias $b_\ell^{XY}$ for
several values of $\ell$, averaged over 2000 lensed CMB simulations calculated with the
precision settings
used in this work (black). In red the same
quantities determined from 400 lensed CMB simulations calculated with original Lenspix interpolation
algorithm with} \verb|interp_factor| = 2. Error bars represent errors on the mean 
estimated from the simulated values.
}
\label{fig:cls_bias}
\end{figure}

\begin{figure}
\center
\includegraphics[width = 0.49 \textwidth]{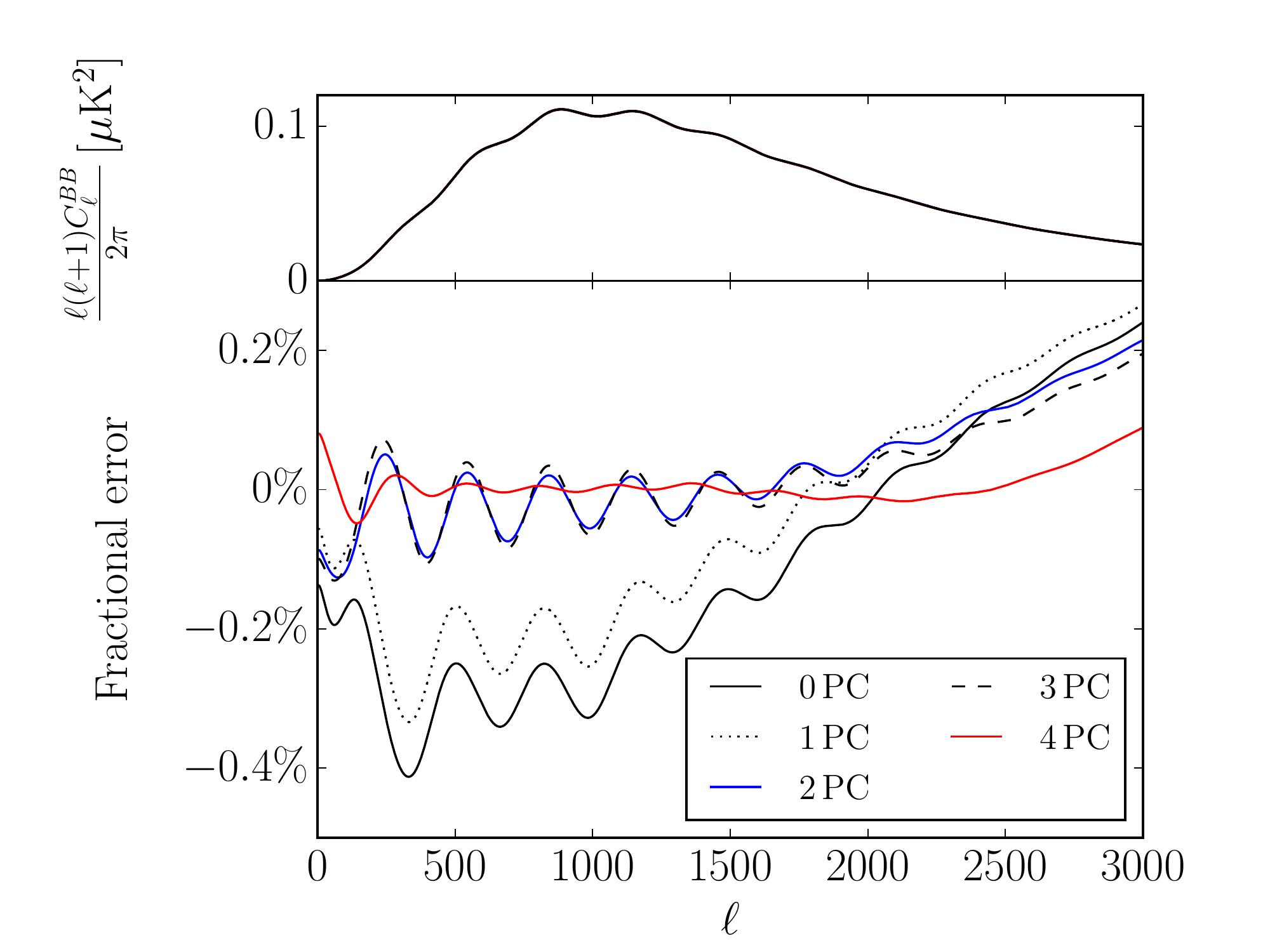}
\caption{Lensed $C_\ell^{BB}$ calculated with lensing potential increased by $\Delta
C^\PP_\ell$ which corresponds to parameter shifts listed in Tab.~\ref{tab:shifts} and its
representation in terms of the first $N \in  {0 \ldots 4}$ PCs (top: absolute; bottom
percent error between the two).  
} \label{fig:npca}
\end{figure}

To judge agreement between the lensed power spectra from simulations and the theoretical
expectation calculated by CAMB, we define bias variables
\be
	b^{XY}_\ell = \frac
	{
	\sum_{\ell'}
	\ell'(\ell'+1)\Delta C_{\ell'}^{XY}
	}
	{
	\sum_{\ell'}
	\ell'(\ell'+1)C_{\ell'}^{XY,\mathrm{fid}}
	} ,
\ee
which can be evaluated for each simulated CMB sky.
Here the sums go over a bin of width $\Delta \ell = 200$ centered on $\ell$ and 
$\Delta C_{\ell'}^{XY}$ is a difference between simulated and expected value of power
spectra, defined in \eqref{DeltaC}. In Fig.~\ref{fig:cls_bias} we plot average values
of $b^{XY}_\ell$ from simulations for several values of $\ell$; we show the levels of
bias achieved with both the simulations settings used in this work and the original 
Lenspix code with \verb|interp_fact| = 2. In the latter, a small bias is
visible for $XY = TT, EE$.  

\section{Sky Coverage and Optical Depth}
\label{sec:no_low_ell}

A ground-based CMB Stage 4 experiment is unlikely to usefully measure CMB temperature and polarization on the full sky. For that
reason, in this Appendix we use  a Fisher analysis to reexamine some of the results of 
\S\ref{sec:cosmological_parameters}
 for an experiment which observes 40\% of the sky and measures
temperature and
polarization power spectra in the multipole range $\ell = 30 - 3000$. We additionally neglect covariance induced by the sky mask by simply scaling the full sky covariance with the sky fraction. Information from the 
largest scales is
represented by adding a Planck-like prior on $\tau$, corresponding to a standard deviation
of $\sigma_\tau = 0.01$. 

As is to be expected, absence of the large scale measurements
significantly degrades the absolute constraints on  cosmological parameters. However, the relative effects of
the non-Gaussian covariance do not become significantly more important.  
For example, degradation of the most affected parameter
combination, as expressed through the ratio $\sigma_M^\mathrm{ng}/\sigma_M^\mathrm{g}$ in
$\Lambda$CDM and analogous ratios for the other cosmological models, increases by
less than 4\% when omitting information from $\ell < 30$.
The largest effect of this omission is in the $\tau-A_s$ plane.
When {large angle polarization data are improved over Planck, they {further} break the $A_s e^{-2\tau}$ degeneracy in the heights of the acoustic peaks. Without this improvement,  lensing measurements become {more competitive} in breaking} this degeneracy and consequently constraints on these two parameters
are degraded by $\sim 10\%$ in  all
three cosmological models investigated in \S\ref{sec:cosmological_parameters}.

\begin{figure}
\center
\includegraphics[width = 0.49 \textwidth]{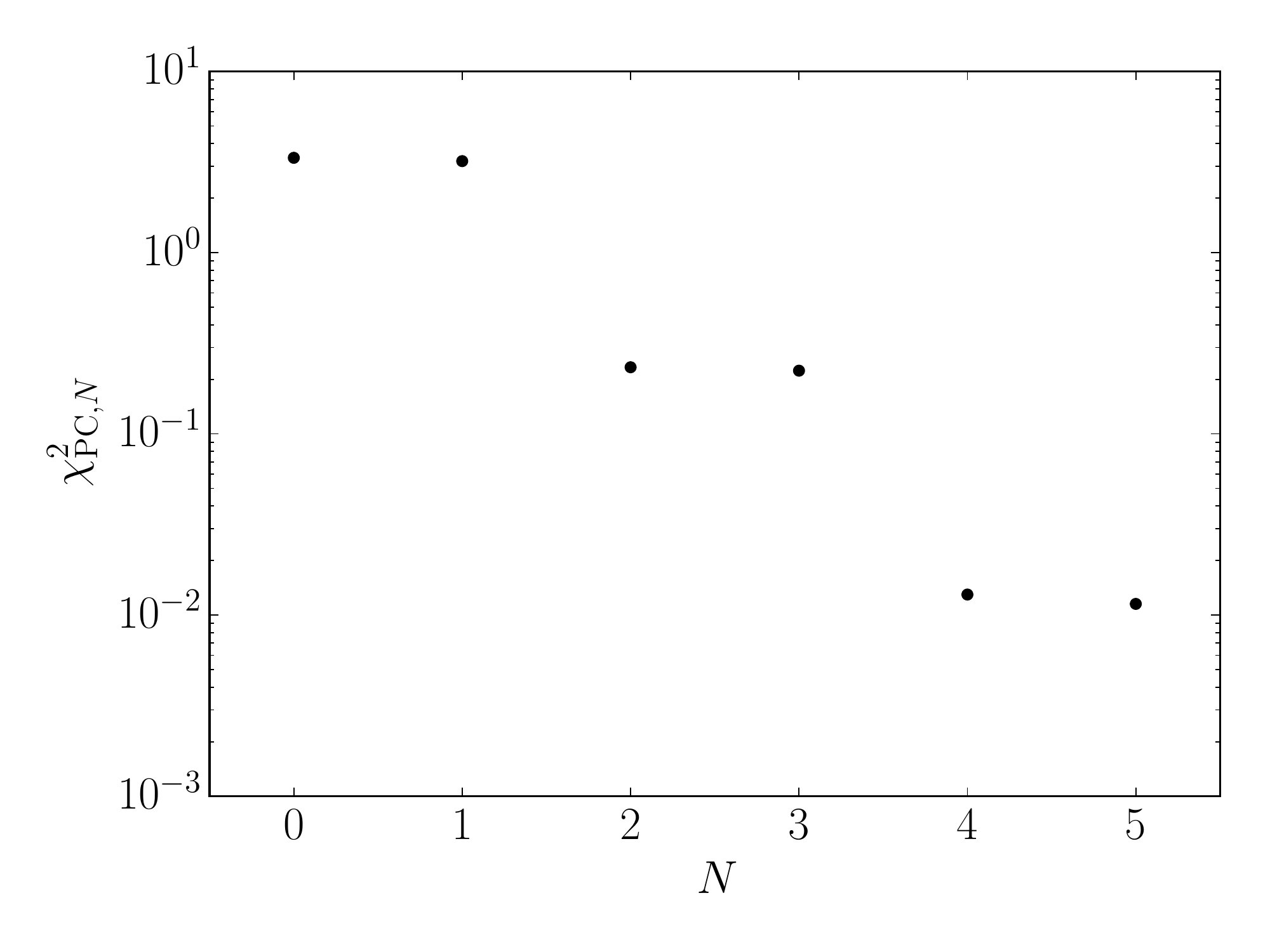}
\caption{Dependence of $\chi^2_{\mathrm{PC},N}$ \eqref{gamma_n}, measure of error caused by
approximating the lensing potential using first $N$ principal component, on $N$.
In this case $\Delta C_\ell^\PP$ corresponds to parameter shifts listed in Tab.~\ref{tab:shifts}.}
\label{fig:pcachi2}
\end{figure}

\section{Number of lensing principal components}
\label{sec:n_pca}

Two considerations guide the choice of the number of principal components of the lensing potential to be measured from the lensed CMB power spectra. Keeping a larger number
of PCs leads to a more accurate description of the lensed power spectra. On the other hand,
increasing number of parameters slows down convergence of the MCMC calculations, requiring physicality priors, since the higher PCs are more poorly constrained by definition. In this section we justify our choice of
using five PCs.

First we look at the fidelity in reproducing lensing effects in the observed
 $C^{XY}_\ell$ power spectra.
For definiteness, we perturb the {fiducial} $C^\PP_\ell$ by  a $\Delta C^\PP_\ell$ which corresponds to shifts in the cosmological parameters given in Table~\ref{tab:shifts} (at fixed
unlensed power spectra). This
change represents a realistic change in the lensing potential which might be encountered
in a real analysis, as cosmological model with parameters from Tab.~\ref{tab:shifts} is
between 68\% and 95\% probability contours for $\Lambda$CDM+$w$ model in the simulation
investigated in the main text.
For completeness and comparison to the results in the main text, the lensing
principal components arising from this change are 
\be
\label{appB_theta}
	\Theta^{(i)}= \{ -0.012,   0.11,   -0.011,  -0.06,  -0.012\} .
\ee
Note that the investigated parameter change is not aligned with change of $M^w$ from
\eqref{appB_theta}, nor need it be since it represents the degeneracy direction rather than the
direction most affected by lens covariance.

\begin{table}
\caption{Shifts in the cosmological parameters used to probe  
approximations of the lensing potential in terms of lensing PCs}
\label{tab:shifts}
\begin{tabular}{c|c}
\hline\hline
Parameter & Shift \\
\hline
$h$ & 0.175\\
$\Omega_c h^2$ & $-1.0 \times 10^{-3}$\\
$\Omega_b h^2$ & $3.5\times 10^{-5}$\\
$n_s$ & $1.2 \times 10^{-3}$\\
$A_s$ & $-1.6 \times 10^{-11}$ \\
$\tau$ & $-2.2 \times 10^{-3}$\\
$w$ & $-0.52$ \\
\hline\hline
\end{tabular}
\end{table}

In top panel of Fig.~\ref{fig:npca} we show the resulting $C^{BB}_\ell$ power
spectrum calculated with lensing potential changed by the full $\Delta C^\PP_\ell$
vs.~when this change is approximated using the first $N \in
\{0, 1,2,3,4\}$ lensing PCs; the difference is too small to be visible directly and so the bottom
panel shows the percent error.
Note that the smallness of these changes explains why in the main text
approximations based on
linearizing power spectra deviations and PC amplitudes are excellent even for relatively
large cosmological parameter shifts.

Since the change in Eq.~\ref{appB_theta} is dominated by $\Theta^{(2)}$, most of the 
improvement in fidelity comes when adding that component.
  In fact for this particular $\Delta
C^\PP_\ell$, first two PCs are sufficient to faithfully describe effects of lensing in
$TT$, $EE$ and $TE$ power spectra extremely well.  The next large jump in fidelity comes with the  fourth PC which is associated with the high multipole range of $BB$
 in Fig.~\ref{fig:npca}.  We checked several other choices of
allowed $\Delta C^{\PP}_\ell$ and for all of them four principal components lead to
small errors on the power spectra level.

To quantify the total significance of the errors we construct
 \be
\label{gamma_n}
	\chi^2_{\mathrm{PC},N} = \sum_{
		\substack{
			\text{$ i,j$}
		}}
	\delta D_i(N)
	\(\Cov_{i,j}\)^{-1}
	\delta  D_j(N) ,
\ee
where for brevity we introduce $\delta D_i(N)=\delta
C_\ell^{XY}(N)$, the power spectrum error caused by  approximating $\Delta C^\PP_\ell$ using the first $N$ PCs, with
$i$ indexing all multipoles and power spectra types.
As $N$ increases, the PCs approximate the full effect of lensing better and
$\chi^2_{\mathrm{PC},N}$ decreases. In Fig.~\ref{fig:pcachi2} we show this dependence; as we saw before,
adding fourth PC leads to a significant improvement in our ability to capture the effects
of gravitational lensing on the CMB. 
For some choices of $\Delta C^\PP_\ell$, adding fifth PC improves 
$\chi^2_{\mathrm{PC},N}$ by a factor of a few on top of the $\sim$ hundred-fold
improvement in $\chi^2_{\mathrm{PC},N}$ arising from using four PCs. For this work we
decided to include fifth PC into the analysis as well, even though its inclusion is not
expected to have any significant impact on the results.

\bibliography{lenspc}

\end{document}